\documentclass{statsoc}
\usepackage[a4paper,top=1in, bottom=1in, left=0.95in, right=0.95in]{geometry}
\usepackage{amsmath, amssymb}
\usepackage{arydshln}
\usepackage{enumerate}
\usepackage{setspace}
\usepackage{natbib}
\usepackage{hyperref}
\usepackage{mathtools}
\usepackage{xcolor}
\hypersetup{
  colorlinks=true,
  citecolor=black,
  linkcolor=black}
\usepackage{booktabs}
\usepackage{algorithm}
\usepackage{algpseudocode}
\usepackage{graphicx}
\usepackage{pifont}
\newcommand{\cmark}{\ding{51}}%
\newcommand{\xmark}{\ding{55}}%

\makeatletter
\renewcommand*{\eqref}[1]{%
  \hyperref[{#1}]{\textup{\tagform@{\ref*{#1}}}}%
}
\makeatother

\newcommand{\const}{B}
\newcommand{\lb}{\left(}
\newcommand{\rb}{\right)}
\newcommand{\td}{\tilde}
\newcommand{\E}{\mathbb{E}}
\newcommand{\cE}{\mathcal{E}}
\newcommand{\cC}{\mathcal{C}}

\newcommand{\R}{\mathbb{R}}
\renewcommand{\r}{r}

\renewcommand{\P}{\mathbb{P}}

\newcommand{\A}{\mathcal{A}}
\newcommand{\D}{\mathcal{D}}
\newcommand{\Z}{\mathcal{Z}}
\newcommand{\I}{\mathcal{I}}
\newcommand{\eps}{\epsilon}
\newcommand{\qt}{\mathrm{Quantile}}
\newcommand{\lo}{\mathrm{lo}}
\newcommand{\hi}{\mathrm{hi}}
\newcommand{\obs}{\mathrm{obs}}
\newcommand{\Var}{\mathrm{Var}}
\newcommand{\tr}{\mathrm{tr}}
\newcommand{\ca}{\mathrm{ca}}

\newcommand{\tv}{d_{\mathrm{TV}}}

\newcommand{\CITE}{\hat{C}_\text{ITE}}
\newcommand{\tdCITE}{\td{C}_\text{ITE}}
\newcommand\indep{\protect\mathpalette{\protect\independenT}{\perp}}
\def\independenT#1#2{\mathrel{\rlap{$#1#2$}\mkern2mu{#1#2}}}

\DeclarePairedDelimiterX{\divpq}[2]{(}{)}{%
  #1\;\delimsize\|\;#2%
}
\allowdisplaybreaks

\numberwithin{equation}{section}

\newtheorem{theorem}{Theorem}
\newtheorem{proposition}{Proposition}
\newtheorem{lemma}{Lemma}
\newtheorem{remark}{Remark}
\newtheorem{corollary}{Corollary}

\definecolor{darkgreen}{rgb}{0.0, 0.42, 0.24}

\title[Conformal Inference of Counterfactuals and Individual Treatment Effects]{Conformal Inference\\ of Counterfactuals and Individual Treatment Effects}

\author{Lihua Lei}
\address{Department of Statistics, Stanford University}
\email{lihualei@stanford.edu}

\author[Lihua Lei and Emmanuel J. Cand\`{e}s]{Emmanuel J. Cand\`{e}s}
\address{Department of Statistics and Department of Mathematics, Stanford University}
\email{candes@stanford.edu}

\begin{document}

\maketitle

\begin{abstract}
  Evaluating treatment effect heterogeneity widely informs treatment
  decision making. At the moment, much emphasis is placed on the
  estimation of the conditional average treatment effect via flexible
  machine learning algorithms. While these methods enjoy some
  theoretical appeal in terms of consistency and convergence rates,
  they generally perform poorly in terms of uncertainty
  quantification. This is troubling since assessing risk is crucial
  for reliable decision-making in sensitive and uncertain
  environments. In this work, we propose a conformal inference-based
  approach that can produce reliable interval estimates for
  counterfactuals and individual treatment effects under the potential
  outcome framework. For completely randomized or stratified randomized experiments with perfect
  compliance, the intervals have guaranteed average coverage in finite
  samples regardless of the unknown data generating mechanism. For
  randomized experiments with ignorable compliance and general
  observational studies obeying the strong ignorability assumption,
  the intervals satisfy a doubly robust property which states the
  following: the average coverage is approximately controlled if
  either the propensity score or the conditional quantiles of
  potential outcomes can be estimated accurately.  Numerical studies
  on both synthetic and real datasets empirically demonstrate that
  existing methods suffer from a significant coverage deficit even in
  simple models. In contrast, our methods achieve the
  desired coverage with reasonably short intervals. 
\end{abstract}

\section{From Average Effects To Individual
  Effects}\label{sec:introduction}

Estimating the average treatment effect (ATE) for a population of
interest has been a main focus of a rich literature on causal
inference, and the last decades have seen the development of extensive
statistical theory addressing important issues such as identification,
estimation, and uncertainty quantification
\citep[e.g.][]{rubin1974estimating, pearl1995causal}. That said, the
average effect only provides a coarse summary of the distribution of a
treatment effect, and may be insufficient, or even misleading, to
validate an intervention. Imagine, for instance, that a drug cures
70\% of the patients while it makes the symptoms much worse for the remaing
30\%. It is unclear whether the drug should be approved in spite of the
positive average effect. This is not an artificial example; 
trials typically do not give clear cut results and examples of this kind are common.
On May 31, 2018, the National Academy of Medicine (NAM) held a
workshop to discuss approaches of examining individual treatment
effects (ITE) to support individualized patient care. In conjunction
with this event,  NAM published a report highlighting the importance
of ITE in medicine.  We quote from this document:
\begin{quote}
``The individuality of the patient should be at the core of every treatment decision. One-size-fits-all approaches to treating medical conditions are inadequate; instead, treatments should be tailored to individuals based on heterogeneity of clinical characteristics and their personal preferences.''
\end{quote}
Outside of medical science, treatment effect heterogeneity is also of
great concern to political scientists \citep{imai2011estimation,
  grimmer2017estimating}, psychologists \citep{bolger2019causal,
  winkelbeiner2019evaluation}, sociologists \citep{xie2012estimating,
  breen2015heterogeneous}, economists
\citep{florens2008identification, djebbari2008heterogeneous}, and
education researchers \citep{morgan2001counterfactuals,
  brand2010benefits}. In short, there is a wide range of fields that
would benefit from a better understanding of ITE. 



To move past the average treatment effect as the object of inference,
most existing works have targeted, instead, the {\em conditional
  average treatment effects} (CATE)---although we will introduce a
formal definition later in Section \ref{sec:setup}, this is the
expectation of the ITE conditional on the values of the covariates. 
While CATE naturally provide a richer summary than ATE,
they surely still neglect the inherent variability in the response
(the conditional variance, if you will) which might be crucial for
decision-making; unless, of course, the covariates explain away most
of the variation in ITE. 
As a consequence, two types of variability are of immediate concern:
(1) the variability of the response around the regression function
and, (2) the variability of CATE estimators due to finite
samples. Getting both (1) and (2) under control requires having
sufficiently many covariates to explain away a significant fraction of
the variability in the response, and at the same time, a nearly
perfect estimate of CATE for every value of the covariates. Neither of
these seem to be realistic for everyday causal inference problems.

  In areas like medical science or public policy, point estimates are
  insufficient to inform decisions due to the huge loss potentially
  incurred by wrong actions. A confidence interval, or at least a
  p-value, is required by the U.S.~Food and Drug Administration to
  approve a drug, in order to guarantee sufficient evidence and
  confidence in favor of the drug. The issues are that despite the
  importance of uncertainty quantification, the reliability of modern
  machine learning methods is typically under-studied, and that
  theoretical guarantees are hard to come by. For instance, existing
  theory usually requires strong non-verifiable assumptions and
  asymptotic regimes one does not encounter in practice. This of
  course limits the applications of machine learning methods in
  sensitive causal inference problems.  In addition, we will
  demonstrate later that the confidence intervals for CATE or prediction intervals for ITE
  produced by frequently discussed methods (including Bayesian
  methods) typically have unsatisfactory or unacceptable coverage,
  even in very simple and smooth models with at most ten
  covariates.


In this work, we leverage ideas from conformal inference to construct
valid prediction intervals for individual treatment effects under the
potential outcome framework \citep{neyman1923application,
  rubin1974estimating}. In particular, we address two challenges:
\begin{enumerate}[(1)]
\item Construct intervals for ITE with reliable coverage for subjects
  in the study, for which one of the potential oucomes is missing;
\item Construct intervals for ITE with reliable coverage for subjects
  not in the study, and for which both potential oucomes are missing.
\end{enumerate}
To be sure, in the case where a method is able to estimate ITE
accurately, there is no need to distinguish these two tasks. In
practice, however, this is rarely possible due to insufficient sample
sizes, model misspecification, and inherent variability. We will thus see that the second challenge is generally far more ambitious since two potential outcomes are never observed simultaneously; consequently we cannot model them jointly.

In brief, Section \ref{sec:counterfactual} shows how the first
challenge reduces to counterfactual inference since one potential
outcome is observed for each subject; here, an interval for the
missing outcome can be shifted into an interval for the individual
treatment effect by contrasting it with the observed outcome.  This
section also introduces methods with guaranteed coverage even under
model misspecification. In particular, for completely randomized or
stratified randomized trials with perfect compliance so that the
propensity score is known but may not be constant, our method
achieves coverage in finite samples \emph{without any assumption other
  than that of operating on ~i.i.d.~samples}. 
For general observational studies under the strong
ignorability assumption \citep{rubin1974estimating}, or randomized
experiments with ignorable compliance, our methods have guaranteed
coverage provided that either the outcome model or the treatment model
is accurately estimated. This is analogous to the \emph{doubly robust}
property applicable to average treatment effect
\citep[e.g.][]{robins1994estimation, kang2007demystifying} which
speaks to the consistency of point
estimates. 

Having addressed the first challenge, we will see in Section
\ref{sec:effect} that our methods can serve as a stepping stone for
the second. A naive approach, here, would be to apply the
counterfactual inference on both potential outcomes and contrast the
two intervals to induce an interval for ITE.
Moreover, we introduce another approach 
which applies the counterfactual inference to generate intervals for ITE
applicable to subjects in the dataset as an intermediate step, and
trains a model to generalize these intervals to subjects not in the
study.

Finally, our methods easily extend to a widely studied problem which
goes by the name of generalizability, or transportability, or external
validity, and which concerns settings in which there is a
distributional shift between the target population and the study
population \citep[e.g.][]{stuart2011use, tipton2014generalizable}, see
Sections \ref{sec:setup}--\ref{sec:effect}. Furthermore, as explained
in Section \ref{sec:extension}, our methods also naturally adapt to
other causal inference frameworks, such as causal diagrams
\citep{pearl1995causal} and invariant prediction
\citep{peters2016causal}.


\section{From Point Estimates To Interval Estimates}
\label{sec:setup}
\subsection{Problem setup}
Throughout the paper we focus on the potential outcome framework
\citep{neyman1923application, rubin1974estimating} with a binary
treatment. Extensions to other causal inference frameworks are
discussed in Section \ref{sec:extension}. Given $n$ subjects, denote
by $T_{i}\in \{0, 1\}$ the binary treatment indicator, by
$(Y_{i}(1), Y_{i}(0))$ the pair of potential outcomes and by $X_{i}$
the vector of other covariates. We assume that
\[(Y_{i}(1), Y_{i}(0), T_{i}, X_{i})\stackrel{i.i.d.}{\sim}(Y(1), Y(0), T, X),\]
where $(Y(1), Y(0), T, X)$ denotes a generic random vector. Under the
{\em stable unit treatment value assumption} (SUTVA) commonly assumed
in the literature \citep{rubin1990formal}, the observed dataset comprises triples $(Y_{i}^{\obs}, T_{i}, X_{i})$ where
\[Y_{i}^{\obs} = \begin{cases} Y_{i}(1), & T_{i} = 1,\\
 Y_{i}(0), & T_i = 0.\end{cases} 
\]
The individual treatment effect $\tau_{i}$ is defined as
\begin{equation}
\label{eq:ITE}
\tau_{i} \triangleq  Y_{i}(1) - Y_{i}(0).
\end{equation}
By definition, only one potential outcome is observed for every unit while the other is missing. Therefore, the ITE are unobserved and have to be inferred. Throughout the paper we assume the strong ignorability:
\begin{equation}
  \label{eq:ignorability}
  (Y(1), Y(0)) \indep T \mid X.
\end{equation}
Strong ignorability rules out any source of unmeasured confounders,
which affect both the treatment assignment and the potential
outcomes. Under this assumption, the treatment assignment is purely
randomized conditional on any covariate values. Although ignorability
is a strong assumption, it is a widely used starting point to
formulate statistical theory and methodology \citep{rubin1978bayesian, rosenbaum1983central, imbens2015causal}. 

\subsection{Traditional inferential targets}
As mentioned earlier, existing methods mostly focus on CATE, defined
as
\begin{equation}
  \label{eq:CATE}
  \tau(x) \triangleq \E [Y(1) - Y(0) \mid X = x].
\end{equation}
The CATE function $\tau(\cdot)$ can be expressed as
$\tau(x) = m_{1}(x) - m_{0}(x)$, where
\[
m_{1}(x) = \E [Y(1) \mid X = x] \, \text{ and } \,  m_{0}(x) = \E [Y(0) \mid X =
x].
\]
It is standard in the literature to impose modeling restrictions on
$m_{1}(x)$ and $m_{0}(x)$ (or equivalently $\tau(x)$ and $m_{0}(x)$)
and estimate these functions using parametric or nonparametric
techniques. Under the strong ignorability assumption, a standard
argument shows that
\[m_{1}(x) = \E [Y^{\obs} \mid X = x, T = 1] \, \text{ and } \,
  m_{0}(x) = \E [Y^{\obs} \mid X = x, T = 0].\] Thus, the estimation
problem reduces to that of estimating certain conditional
  expectations \citep[e.g.][]{nie2017quasi, foster2019orthogonal,
    kennedy2020optimal}. However, drawing reliable confidence
bands---which can be trusted in practice---from finite samples around
estimates of possibly high-dimensional regression functions
is delicate,
to say the least.  As we will show later, many standard techniques,
e.g. normal approximation and resampling techniques, may significantly
underestimate the variability of such estimates.

An alternative estimand is the conditional quantile treatment effect
(CQTE), which happens to be less investigated in the literature
\citep[e.g.][]{koenker1978regression, fort2016unconditional}. Instead of contrasting the mean
functions $m_{1}(x)$ and $m_{0}(x)$, 
CQTE is defined as the difference between the $\beta$-th quantiles of
the distributions of $Y(1)$ and $Y(0)$ conditional on $X = x$ for a
given $\beta$ of interest. 
CQTE is to be distinguished from the quantiles of
$Y(1) - Y(0)$; they are not the same at all!
It turns out that the conditional quantiles of ITE are unidentifiable
in general since they involve the joint distribution of $(Y(0), Y(1))$
and that we can never observe joint outcomes.  Furthermore, the
difficulties associated with uncertainty quantification
persist.
 
\subsection{Coverage of interval estimates}
In this work, we take counterfactuals $(Y_{i}(1), Y_{i}(0))$'s and the
ITE $\tau_{i}$'s as objects of inference, and attempt to construct
prediction intervals covering these random variables. Taking the
potential outcome $Y(1)$ as an example, we wish to construct intervals
$\hat{C}_{1}(x)$, which depend on the location in covariate space,  and
obey
\begin{equation}
  \label{eq:coverage}
  \P(Y(1) \in \hat{C}_{1}(X))\ge 1 - \alpha,
\end{equation}
for a pre-specified level $\alpha$. 
Similarly, we seek $\hat{C}_{0}(x)$ for $Y(0)$ and $\CITE(x)$
for $Y(1) - Y(0)$ obeying marginal coverage in the same sense, i.e.
\begin{equation}
  \label{eq:covITE}
   \P(Y(1) - Y(0) \in \CITE(X))\ge 1 - \alpha. 
\end{equation}

If we had perfect knowledge of the the quantiles $q_{\beta}(x)$ of
$Y(1)$ given $X = x$ for each $\beta\in (0, 1)$, then the oracle
estimate, 
\[
{C}_{1}(x) = [q_{\alpha / 2}(x), q_{1 - \alpha / 2}(x)],
\]
would automatically satisfy \eqref{eq:coverage}. This is arguably the
best prediction interval one could produce. In addition, coverage
would hold conditionally on $X = x$. In reality, conditional quantiles may
be hard to estimate due to the limited effective sample size and
imperfect model knowledge. In this case, substituting the true
quantiles in $C_1(x)$ with estimates may fail to yield valid coverage.

A typical objection to the criterion \eqref{eq:coverage} is that it
only controls coverage in an average (marginal) sense---just as the
root mean squared error (RMSE) measures {\em average}
performance. Admittedly, it does not say much about the validity of
the predicted range for a patient with {\em this} $x$.
Without modeling assumptions, it is known to be impossible to
construct non-trivial prediction intervals with {\em guaranteed}
conditional coverage \citep{barber2019limits}. This does not mean that
conditional coverage cannot be achieved in any particular application;
in fact, we make conditional coverage a focus point of this work and
demonstrate reasonable approximations.


\subsection{General coverage criteria}\label{subsec:general_criterion}
In classical causal inference, it is often argued that the average
treatment effect on the treated (ATT), defined as
$\E [Y(1) - Y(0)\mid T = 1]$, is preferrable to the average
treatment effect (ATE), defined as $\E [Y(1) - Y(0)]$, 
because it is often more plausible to remove treatment from treated units than to assign treatment to control units. Theoretically, the identification of ATT is strictly easier than ATE since the former requires a weaker assumption on the propensity scores \citep[e.g.][]{imbens2015causal}. 
The criterion \eqref{eq:coverage} can be then modified as
\begin{equation}
  \label{eq:coverage_ATT}
  \P(Y(t) \in \hat{C}_{t}(X) \mid T = 1)\ge 1 - \alpha, \quad (t = 0, 1).
\end{equation}
Compared to \eqref{eq:coverage}, \eqref{eq:coverage_ATT} substitutes
the marginal distribution of $X$ with the conditional distribution of
$X$ given $T = 1$ (recall that we operate under the strong
ignorability assumption). Similarly we may consider the counterpart of
the average treatment effect on the controls (ATC) by conditioning on
$T = 0$. These considerations motivate the following general
criterion:
\begin{equation}
  \label{eq:coverage_general}
    \P_{(X, Y(t))\sim Q_{X}\times P_{Y(t)\mid X}}(Y(t) \in \hat{C}_{t}(X))\ge 1 - \alpha, \quad (t = 0, 1).
  \end{equation}
  For instance, \eqref{eq:coverage_ATT} is a special case with
  $Q_{X} = P_{X\mid T = 1}$. The general formulation
  \eqref{eq:coverage_general} is useful when the study population
  differs from the target population. In this case, $Q_{X}$ can be
  chosen to be the covariate distribution in the target
  population. Inference on ATE and CATE in this setting has been a
  subject of much recent research, and is known under the name of
  generalizability, or transportability, or external validity
  \citep[e.g.][]{tipton2013improving, pearl2014external}.


\section{From Observables To Counterfactuals}\label{sec:counterfactual}

\subsection{Counterfactuals and covariate shift}\label{subsec:counterfactuals_covariate_shift}
Counterfactual inference is both the ultimate goal in areas such as
policy evaluation \citep[e.g.][]{athey2018matrix, ben2018augmented,
  arkhangelsky2019synthetic} as well as the stepping stone for
inferring ITE in general. We construct prediction intervals for $Y(1)$
and $Y(0)$ using the i.i.d.~observations
$(Y_{i}^{\obs}, T_{i}, X_{i})$, and make no assumption other than the
strong ignorability assumption. Therefore, only the samples in the
treatment (resp.~control) group are useful for constructing
$\hat{C}_{1}(x)$ (resp.~$\hat{C}_{0}(x)$).

Under SUTVA and strong ignorability assumption, the joint distribution of
$(X, Y^{\obs})$ of the observed treated samples is given by 
\[
P_{X\mid T = 1} \times P_{Y(1) \mid X}.
\]
Once again, we want to achieve \eqref{eq:coverage_general} under the
target distribution $Q_{X}\times P_{Y(1)\mid X}$ on the basis of an
i.i.d.~sample drawn from $P_{X\mid T = 1}\times P_{Y(1) \mid X}$.
These two distributions share the same conditional distribution
$P_{Y(1)\mid X}$ of the outcome but otherwise differ in the
distribution of the covariates. Covariate shifts have been widely
studied in the machine learning literature
\citep[e.g.][]{shimodaira2000improving}, yet, the heavy focus there is
on point estimates. For interval estimates, we shall rely on weighted
conformal inference recently developed by \cite{barber2019conformal}.

\subsection{Weighted conformal inference}
Conformal inference was introduced by Vladimir Vovk and his
collaborators \citep[e.g.][]{vovk2005algorithmic,
  gammerman2007hedging, shafer2008tutorial, vovk2009line,
  vovk2012conditional, vovk2013transductive,
  balasubramanian2014conformal, vovk2015cross}. It later gained
significant attention from the statistics community for regression
problems \citep[e.g.][]{lei2013distribution, lei2014distribution,
  lei2018distribution, barber2019limits, barber2019predictive} and
classification problems \citep[e.g.][]{sadinle2019least,
  romano2020classification}, spurring further developments. Given
i.i.d. samples $(X_{i}, Y_{i})_{i=1}^{n}$ drawn from a distribution
$P_{X}\times P_{Y\mid X}$, conformal inference takes an arbitrary
prediction model---such as an estimate of the conditional
quantile---as input and calibrates it to produce a prediction set
$\hat{C}(x)$ with guaranteed marginal coverage, such that 
\begin{equation}
  \label{eq:conformal_guarantee}
  \P_{(X, Y)\sim P_{X}\times P_{Y\mid X}}(Y \in \hat{C}(X))\ge 1 - \alpha.
\end{equation}
For instance, with estimates $\hat{q}_{\alpha_{\lo}}(x)$ and
$\hat{q}_{\alpha_{\hi}}(x)$ of the $\alpha_{\lo}$-th and
$\alpha_{\hi}$-th conditional quantiles of $Y \mid X = x$, the
conformal quantile regression (CQR) introduced in
\cite{romano2019conformalized} as a variant of the standard conformal
inference, would produce an interval estimate of the form
\begin{equation}
  \label{eq:CQR_CX}
  \hat{C}(x) = [\hat{q}_{\alpha_{\lo}}(x) - \eta,
  \hat{q}_{\alpha_{\hi}}(x) + \eta]; 
\end{equation}
above, $\eta$ is a data-driven constant computed in a particular
way. Conditional quantiles can be estimated via quantile regression \citep[e.g.][]{koenker1978regression, koenker1994confidence, koenker2001quantile, yu2001bayesian, koenker2005quantile, meinshausen2006quantile, koenker2017quantile}. In contrast to asymptopia---we refer here to classical
  asymptotic normality theory or asymptotic empricial process theory
  which characterizes the accuracy of certain resampling
  methods---CQR enjoys finite sample coverage guarantees without regard to the
unknown joint distribution of $X$ and $Y$.

Because our inference problem involves a potential covariate shift
between the target distribution and the sampling distribution, we need
to adjust the criterion \eqref{eq:conformal_guarantee} into
\begin{equation}
  \label{eq:weighted_conformal_guarantee}
  \P_{(X, Y)\sim Q_{X}\times P_{Y\mid X}}(Y \in \hat{C}(X))\ge 1 - \alpha.
\end{equation}
To address such a situation, \cite{barber2019conformal} introduced a
weighted variant of conformal inference achieving
\eqref{eq:weighted_conformal_guarantee}. This holds with the proviso
that the the likelihood ratio $w(x) = dQ_{X}(x) / dP_{X}(x)$ is known,
although the algorithm is shown to perform well when it is only
estimated.  When applied to CQR, the weighted interval estimate is
still of the form \eqref{eq:CQR_CX}, the difference being that the
algorithm computing $\eta(x)$ incorporates information about the
likelihood ratio $w(x)$.  Algorithm \ref{algo:split-CQR} sketches
split-CQR, a type of weighted conformal inference we shall use in this
work.\footnote{\cite{barber2019conformal} introduced weighted
  conformal inference and applied to conditional mean estimates
  \citep[e.g.][]{lei2018distribution}. The extension to other conformal
  inference techniques such as CQR is straightforward.}  In passing,
it is worth mentioning that in this algorithm, $\eta(x)$ remains
invariant if $w(x)$ is rescaled to become $c \cdot w(x)$ in which $c$
is an arbitrary positive constant.

\begin{algorithm}[H]
  \caption{Weighted split-CQR}  \label{algo:split-CQR}
  \textbf{Input: } level $\alpha$, data $\Z = (X_{i}, Y_{i})_{i\in \I}$, testing point $x$, function $\hat{q}_{\beta}(x; \D)$ to fit $\beta$-th conditional 

  \hspace{1.4cm} quantile and function $\hat{w}(x; \D)$ to fit the weight function at $x$ using $\D$ as data

  \textbf{Procedure: }
  \begin{algorithmic}[1]
    \State Split $\Z$ into a training fold $\Z_{\tr} \triangleq (X_{i}, Y_{i})_{i\in \I_{\tr}}$ and a calibration fold $\Z_{\ca}\triangleq (X_{i}, Y_{i})_{i\in \I_{\ca}}$
    \State For each $i \in \I_{\ca}$, compute the score $V_{i} = \max\{\hat{q}_{\alpha_{\lo}}(X_{i}; \Z_{\tr}) - Y_{i}, Y_{i} - \hat{q}_{\alpha_{\hi}}(X_{i}; \Z_{\tr})\}$
    \State For each $i \in \I_{\ca}$, compute the weight $W_{i} = \hat{w}(X_{i}; \Z_{\tr}) \in [0, \infty)$
    \State Compute the normalized weights $\hat{p}_{i}(x) = \frac{W_{i}}{\sum_{i\in \I_{\ca}}W_{i} + \hat{w}(x; \Z_{\tr})}$ and $\hat{p}_{\infty}(x) = \frac{\hat{w}(x; \Z_{\tr})}{\sum_{i\in \I_{\ca}}W_{i} + \hat{w}(x; \Z_{\tr})}$
    \State Compute $\eta(x)$ as the $(1 - \alpha)$-th quantile of the distribution $\sum_{i\in \I_{\ca}}\hat{p}_{i}(x)\delta_{V_{i}} + \hat{p}_{\infty}(x)\delta_{\infty}$
  \end{algorithmic}

  \textbf{Output: $\hat{C}(x) = [\hat{q}_{\alpha_{\lo}}(x; \Z_{\tr}) - \eta(x), \hat{q}_{\alpha_{\hi}}(x; \Z_{\tr}) + \eta(x)]$}  
\end{algorithm}

Note that in step 4, if $\hat{w}(x; \Z_{\tr}) = \infty$, we
  set $\hat{p}_{i}(x) = 0\,\, (i\in [n])$ and
  $\hat{p}_{\infty}(x) = 1$, in which case step 5 gives
  $\hat{C}(x) = (-\infty, \infty)$. The requirement that
  $W_i\in [0, \infty)$ is natural because $X_{i}\sim P_{X}$ and
  $w(X)\in [0, \infty)$ almost surely under $P_{X}$ even if $Q_{X}$ is
  not absolutely continuous with respect to $P_{X}$. If the
likelihood ratio $w(x)$ is known, \cite{barber2019conformal} prove
that the interval $\hat{C}(x)$ from Algorithm \ref{algo:split-CQR}
achieves \eqref{eq:weighted_conformal_guarantee}.
Moreover, 
we show that the inequality
\eqref{eq:weighted_conformal_guarantee} is almost an equality if the
non-conformity scores have no ties and the covariate shift has a
bounded moment , and further, \eqref{eq:weighted_conformal_guarantee} approximately holds if the covariate shift is unknown but estimated well.
\begin{proposition}\label{prop:equal}
  Consider Algorithm \ref{algo:split-CQR} and assume
  $(X_{i}, Y_{i})\stackrel{i.i.d.}{\sim} P_{X}\times P_{Y \mid
    X}$. First, consider the case where $\hat{w}(\cdot) = w(\cdot)$.
    \begin{enumerate}[(1)]
    \item \eqref{eq:weighted_conformal_guarantee} holds without any
      further assumption.
    \item Further, if the non-conformity scores
      $\{V_{i}: i\in \I_{\ca}\}$ have no ties almost surely, $Q_{X}$
      is absolutely continuous with respect to $P_{X}$, and
      $(\E[\hat{w}(X)^{\r}])^{1/\r} \le M_{\r} < \infty$, then
  \[1 - \alpha \le \P_{(X, Y)\sim Q_{X}\times P_{Y \mid X}}(Y \in
    \hat{C}(X))\le 1 - \alpha + c \, n^{1 / \r - 1}.
  \]
  Above, $c$ is a positive constant that only depends on $M_{\r}$ and $\r$.
\end{enumerate}
In the general case where $\hat{w}(\cdot) \neq w(\cdot)$, set
$\Delta_{w} = (1/2)\E_{X\sim P_{X}}|\hat{w}(X) - w(X)|$. Then coverage
is always lower bounded by $1 - \alpha - \Delta_{w}$, and upper
bounded by $1 - \alpha + \Delta_{w} + cn^{1/\r - 1}$ under the same
assumptions as in (2).
  \end{proposition}
The proof is presented in Appendix \ref{app:miscellaneous}. Note that the proposition holds uniformly over all conditional
 distributions $P_{Y\mid X}$ and all procedures used to fit
 conditional quantiles. When $\hat{w}(\cdot) = w(\cdot)$, because we express the dependence on $P_{X}$
 and $Q_{X}$ only through $(\E[w(X)^{\r}])^{1/\r}$, an immediate consequence
 is that when the likelihood ratio is bounded, we can take
 $\r = \infty$ and the upper bound matches the rate for unweighted
 conformal inference. This essentially implies that weighted split-CQR
 has almost exact coverage when the calibration fold is large.


\subsection{The role of the propensity score}
Propensity scores were introduced by \cite{rosenbaum1983central} for
the analysis of observational studies. Given a binary treatment, the
propensity score $e(x)$ is defined as the probability of getting
treated given the covariate value, i.e.
\[e(x) = \P(T = 1\mid X = x).\]
With the full knowledge of $e(x)$, one can identify ATE/ATT/ATC
using inverse propensity weighting (IPW)
\citep{imbens2015causal} without any assumption on potential outcomes
other than some mild moment conditions. This holds with the proviso
that certain overlap/positivity conditions hold. Specifically, if the
overlap condition $0 < e(X) < 1$ holds almost surely, ATE can be
identified as follows:
\begin{equation}
  \label{eq:ATE_IPW}
  \E[Y(1) - Y(0)] = \E \left[w_{1}(X)Y^{\obs}I(T = 1) - w_{0}(X)Y^{\obs}I(T = 0)\right],
\end{equation}
where $w_{1}(x) = 1 / e(x)$ and $w_{0}(x) = 1 / (1 - e(x))$.  This
holds under moment conditions on $1 / e(X) , 1 / (1 - e(X))$, $Y(1)$
and $Y(0)$.  The overlap condition guarantees that $w_{1}(x)$ and 
$w_{0}(x)$ stay finite, and is necessary for identification through
\eqref{eq:ATE_IPW}. 
Indeed, if $e(x) = 0$ for a subset of covariate values with non-zero
probability, there will be no treated unit on that stratum and hence
the stratum-wise ATE can never be identified without modeling
assumptions on the potential outcomes. The overlap condition is as
fundamental as the strong ignorability assumption in observational
studies. We refer the readers to \cite{d2017overlap} for an extensive
discussion. Similarly, under the weaker overlap condition $e(X) < 1$
almost surely, ATT can be identified through \eqref{eq:ATE_IPW} with
$w_{1}(x) = 1 / \P(T = 1)$ and $w_{0}(x) = e(x) / (1 - e(x)) \P(T = 1)$
\citep{hirano2003efficient}.

The weights in IPW estimators are essentially calibrating the observed
covariate distribution to the target one. This is similar in spirit to
weighted conformal inference. For ATE-type conformal
inference on $Y(1)$, a simple application of Bayes formula implies
that
\[w_{1}(x) = \frac{dP_{X}(x)}{dP_{X\mid T = 1}(x)} = \frac{\P(T =
  1)}{e(x)}.\]
Now recall that weighted conformal inference is invariant to rescaling
of the likelihood ratio $w_{1}(x) \propto 1/e(x)$, so that we can
simply ignore the numerator, in which case everything reduces to the
inverse propensity score. Similarly, for inference on $Y(0)$,
$w_{0}(x)$ can be chosen as $1 / (1 - e(x))$. The concurrence with
IPW-type estimators is here unsurprising since the reweighting scheme
is motivated by the covariate shift due to treatment selection in both
settings.

For ATT-type conformal inference on $Y(1)$, $Q_{X} = P_{X\mid T = 1}$.
Thus, no weighting is needed and unweighted conformal inference may be
applied. By contrast, the inference on $Y(0)$ still requires
reweighting because $Q_X = P_{X\mid T = 1}\not = P_{X\mid T =
  0}$. Applying Bayes formula,
\[w_{0}(x) = \frac{\P(T = 0)}{\P(T = 1)}\frac{e(x)}{1 - e(x)}.\]
We can therefore choose $w_{0}(x) = e(x) / (1 - e(x))$ per our
previous discussion. This happens to coincide with the weights used by
the IPW estimator for ATT.

In general, if $Q_{X}$ is the covariate distribution in another
population, as in the context of
generalizability/transportability/external validity, then 
\[
w_{1}(x) = \frac{dQ_{X}(x)}{dP_{X\mid T = 1}(x)} =
\frac{dQ_{X}(x)}{dP_{X}(x)}\frac{\P(T = 1)}{e(x)}. 
\]
In this case, we can choose $w_{1}(x) = (dQ_{X}/dP_{X})(x) / e(x)$ to
be the inverse propensity tilted likelihood ratio. Similarly
$w_{0}(x) = (dQ_{X}/dP_{X})(x) / (1 - e(x))$. All these weight
functions are displayed in Table \ref{tab:weight_func}. In sum,
weighted conformal inference depends on propensity scores in the same
way IPW estimation of average causal effects depends on these same
scores.

\begin{table}
  \caption{\label{tab:weight_func}Summary of weight functions for different inferential targets}

  \centering
  \begin{tabular}{c|cccc}
    \toprule
    Inferential type & ATE & ATT & ATC & General\\
    \midrule
    $w_{1}(x)$ & $1 / e(x)$ & $1$ & $(1 - e(x)) / e(x)$ & $(dQ/dP)(x) / e(x)$\\
    $w_{0}(x)$ & $1 / (1 - e(x))$ & $e(x) / (1 - e(x))$ & $1$ & $(dQ/dP)(x) / (1 - e(x))$\\
    \bottomrule
  \end{tabular}
\end{table}

\subsection{Conformalized counterfactual inference is exact for randomized trials}
For randomized trials with perfect compliance, the strong ignorability
assumption is satisfied by randomization and the propensity score is
known since it is designed by researchers. In completely randomized
experiments, $e(\cdot)$ is a constant mapping, in which case weighting
is not required for either $Y(1)$ or $Y(0)$. For general stratified
experiments such as blocking experiments, $e(x)$ could vary with some
of the covariates (e.g. age, gender), and one would apply weighted
conformal inference by using the weight functions from Table
\ref{tab:weight_func}.  Either way, weighted conformal inference
achieves coverage in finite samples (Proposition \ref{prop:equal})
even if our conditional quantile estimates are completely off.
Moreover, taking $Y(1)$ and the ATE-type coverage as an example,
$\chi^{\r}\divpq{Q_{X}}{P_{X}} \le \E [1 / e(X)^{\r}]$. By Proposition
\ref{prop:equal}, our method has almost exact coverage when the
calibration fold is large and $\E [1 / e(X)^{2}] < \infty$. 
On the other hand, Proposition
  \ref{prop:equal} guarantees that the coverage is lower bounded by
  $1 - \alpha$ no matter the status of the overlap condition. Intuitively, this is because the interval $\hat{C}(x) = (-\infty, \infty)$ and hence always covers $Y(1)$ when $e(x) = 0$.

We close this section with a last important bibliographical
comment. As we were putting the finishing touch on this paper, we
became aware of the independent work by
\cite{kivaranovic2020conformal} applying unweighted standard conformal
inference to construct counterfactual intervals for completely
randomized experiments. Clearly, our two papers have a similar
aim. That said, and as mentioned in their Section 2.1, the approach in
\cite{kivaranovic2020conformal} cannot handle stratified experiments
even if the propensity score is known. Hence, the scopes of the two
papers are very different.

\subsection{Conformalized counterfactual inference is doubly robust}\label{subsec:double_robustness}

For observational studies or randomized trials with imperfect
compliance, the propensity score is unknown and needs to be
estimated. Let $\hat{e}(x)$ denote the estimate of $e(x)$. In this
subsection, we will see that the coverage of weighted split-CQR is
approximately guaranteed if either $\hat{e}(x)\approx e(x)$ or
$\hat{q}_{\beta}(x)\approx q_{\beta}(x)$ with
$\beta \in \{\alpha_{\lo}, \alpha_{\hi}\}$.  Before stating a rigorous
result of this fact, we first provide an intuitive justification. On
the one hand, if $\hat{e}(x)\approx e(x)$, our method approximates the
oracle version of weighted split-CQR with the true weights and the
intervals should, therefore, approximately achieve the desired
coverage even if $\hat{q}_{\beta}(x)$ drastically deviates from the
true conditional quantiles. On the other hand, if
$\hat{q}_{\beta}(x) \approx q_{\beta}(x)$, where $q_{\beta}(x)$ is the
$\beta$-th quantile of $Y(1)$ (or $Y(0)$) given $X = x$, then
\[
V_{i}\approx \max\{q_{\alpha_{\lo}}(X_{i}) - Y_{i}(1), Y_{i}(1) -
q_{\alpha_{\hi}}(X_{i})\}.
\]
As a result,
\[
\P(V_{i}\le 0\mid X_{i}) \approx \P(Y_{i}(1) \in
[q_{\alpha_{\lo}}(X_{i}), q_{\alpha_{\hi}}(X_{i})]\mid X_{i}) =
\alpha_{\hi} - \alpha_{\lo}.
\]
If $\alpha_{\hi} - \alpha_{\lo} = 1 - \alpha$, then $0$ is
approximately the $(1 - \alpha)$-th quantile of the $V_{i}$'s. In
Algorithm \ref{algo:split-CQR} we have that $\eta(x)$ is the
$(1 - \alpha)$-th quantile of the random distribution
$\sum_{i\in \I_{\ca}}\hat{p}_{i}(x)\delta_{V_{i}} +
\hat{p}_{\infty}(x)\delta_{\infty}$.
Denote by $G$ the cumulative distribution function (cdf) of this
random distribution. Then
\[G(0) \approx \E [G(0) \mid (X_i)_{i\in \I_{\ca}}] =
\sum_{i\in \I_{\ca}}\hat{p}_{i}(x)\P(V_{i}\le 0 \mid X_{i}) \approx
\sum_{i\in \I_{\ca}}\hat{p}_{i}(x)(1 - \alpha) \approx 1 - \alpha.\]
This says that $0$ is just about the $(1 - \alpha)$-th quantile of
$G$. This implies that $\eta(x) \approx 0$ and thus
$\hat{C}(x) \approx [q_{\alpha_{\lo}}(x), q_{\alpha_{\hi}}(x)]$. By
definition, the coverage in this case is approximately
$\alpha_{\hi} - \alpha_{\lo} = 1 - \alpha$.

The following theorem, whose proof is in Appendix
\ref{app:miscellaneous}, formalizes the above heuristics. 
\begin{theorem}\label{thm:double_robustness_ATE}
  Let $N = |\Z_{\tr}|$ and $n = |\Z_{\ca}|$. Further, let
  $\hat{q}_{\beta, N}(x) = \hat{q}_{\beta, N}(x; \Z_{\tr})$ be an
  estimate of the $\beta$-th conditional quantile $q_{\beta}(x)$ of
  $Y(1)$ given $X = x$, $\hat{e}_{N}(x) = \hat{e}_{N}(x; \Z_{\tr})$ be
  an estimate of $e(x)$, and $\hat{C}_{N, n}(x)$ be the resulting
  interval from Algorithm \ref{algo:split-CQR}. Assume that
  $\E[1 / \hat{e}_{N}(X)\mid \Z_{\tr}] < \infty$ and
  $\E[1 / e(X)] < \infty$. Assume that one of the following holds:
    \begin{enumerate}[\textbf{A}1]
  \item $\displaystyle \lim_{N\rightarrow \infty}\E\bigg|\frac{1}{\hat{e}_{N}(X)} - \frac{1}{e(X)}\bigg| = 0$;
  \item
    \begin{enumerate}[(1)]
    \item $\alpha_{\hi} - \alpha_{\lo} = 1 - \alpha$,
    \item there exists $r, b_{1}, b_{2} > 0$ such that
      $\P(Y(1) = y \mid X = x) \in [b_{1}, b_{2}]$ uniformly over all
      $(x, y)$ with
      $y\in [q_{\alpha_{\lo}}(x) - r, q_{\alpha_{\lo}}(x) + r]\cup
      [q_{\alpha_{\hi}}(x) - r, q_{\alpha_{\hi}}(x) + r]$, 
    \item there exists $\delta > 0$ such that
      \[\limsup_{N\rightarrow \infty}\E\left[\frac{1}{\hat{e}_{N}(X)^{1 + \delta}}\right] < \infty, \quad \lim_{N\rightarrow \infty}\E\left[\frac{H_{N}(X)}{\hat{e}_{N}(X)}\right] = \lim_{N\rightarrow \infty}\E\left[\frac{H_{N}(X)}{e(X)}\right] = 0,\]
      where
      \[\displaystyle H_{N}(x) = \max\{|\hat{q}_{\alpha_{\lo}, N}(x) - q_{\alpha_{\lo}}(x)|, |\hat{q}_{\alpha_{\hi}, N}(x) - q_{\alpha_{\hi}}(x)|\}.\]
     \end{enumerate}
  \end{enumerate}
  Then under SUTVA and the strong ignorability assumption,
  \begin{equation}
    \label{eq:unconditional_coverage}
    \lim_{N, n\rightarrow\infty}\P_{(X, Y(1))\sim P_{X}\times P_{Y(1)\mid X}}(Y(1) \in \hat{C}_{N, n}(X))\ge 1 - \alpha.
  \end{equation}
  Furthermore, if \textbf{A}2 holds, then for any $\eps > 0$,
  \begin{equation}
    \label{eq:conditional_coverage}
    \lim_{N, n\rightarrow\infty}\P_{X\sim P_{X}}\lb \P(Y(1)\in \hat{C}_{N, n}(X)\mid X)\le 1 - \alpha - \eps\rb = 0.
  \end{equation}
\end{theorem}
Theorem \ref{thm:double_robustness_ATE} is a special case of Corollary
\ref{cor:double_robustness} in Appendix \ref{app:proofs} on the double
robustness of general weighted split-CQR. The refined
  theorems in Appendix \ref{app:proofs} also provide the rate of
  convergence with which coverage is achieved. We choose here to
  present a simpler version to avoid mathematical
  complications. Observe that it is a simple exercise to extend
Theorem \ref{thm:double_robustness_ATE} to other types of coverage and
to $Y(0)$ by consulting Table \ref{tab:weight_func}.

Property \eqref{eq:unconditional_coverage} is analogous to the
doubly robust point estimation of ATE
\citep[e.g.][]{robins1994estimation, kang2007demystifying}, which
yields consistent estimators if either the propensity score or the
conditional mean of potential outcomes are consistent. Nonetheless, we
emphasize that our double robustness is not the same since consistency
of point estimates and coverage of interval estimates are different
concepts.

Property \eqref{eq:conditional_coverage} implies that weighted
split-CQR has approximately guaranteed conditional coverage if the
conditional quantiles are estimated accurately. This is of course
sufficient but not necessary. In practice, we may work with less
accurate estimates and shall nevertheless empirically demonstrate the
robustness of weighted split-CQR in terms of conditional coverage.

\subsection{Numerical experiments}\label{subsec:simul}

In this subsection we demonstrate the performance of our methods via
simulation studies. In particular, we consider a variant of the
example in \cite{wager2018estimation}:
\begin{itemize}
\item The covariate vector $X = (X_{1}, \ldots, X_{d})^{T}$ is such
  that $X_{j} = \Phi(X_{j}')$, where $\Phi$ denotes the cdf of the
  standard normal distribution and $(X_{1}', \ldots, X_{d}')$ is an
  equicorrelated multivariate Gaussian vector with mean zero and
  $\Var(X'_j) = 1$, $\operatorname{Cov}(X'_j, X'_{j'}) = \rho$ for $j \neq j'$.   
  When $\rho = 0$, $X$ is uniformly distributed on the unit
  cube. When $\rho > 0$, the variables are positively correlated.
\item The baseline potential outcome is such that $Y(0)\equiv 0$. This
  simplifies the problem into a pure counterfactual inference problem.
  \item The potential outcome $Y(1)$ is generated as follows:
    \[\E[Y(1) \mid X] = f(X_{1})f(X_{2}),\quad  f(x) = \frac{2}{1 + \exp\{-12(x - 0.5)\}},\]
    which is the same as in \cite{wager2018estimation}, and
    \[Y(1) = \E[Y(1) \mid X] + \sigma(X)\eps, \quad \eps \sim N(0,
    1);\]
    the homoscedastic case $\sigma^{2}(x)\equiv \sigma^{2}$ is
    considered in \cite{wager2018estimation};
  \item The propensity score $e(x)$ is set as:
    \[e(x) = \frac{1}{4}\left\{ 1 + \beta_{2, 4}(x_{1})\right\},\]
    where $\beta_{2, 4}$ is the cdf of the beta
    distribution with shape parameters $(2, 4)$. This ensures that
    $e(x)\in [0.25, 0.5]$, thereby providing sufficient overlap.
  \end{itemize}
  In our experiments, we will consider $8 = 2\times 2 \times 2$
  scenarios: low ($d = 10$) and high ($d = 100$) dimensions,
  uncorrelated ($\rho = 0$) and correlated ($\rho = 0.9$) covariates,
  and homoscedastic ($\sigma^{2}(x) \equiv 1$) and heteroscedastic
  $(\sigma^{2}(x) = -\log x_{1})$ errors.

  We present comparisons with three competing methods offering
  qualitatively different approaches to uncertainty quantification as
  well as well-written \texttt{R} packages:
  \begin{itemize}
  \item Causal Forest \citep{wager2018estimation} uses the
    infinitesimal jackknife \citep{efron2014estimation,
      wager2014confidence} to estimate the variance of CATE
    estimators. The authors established asymptotically valid coverage
    of CATE under regularity assumptions. This work does not
    discuss ITE. The method is implemented in the \texttt{grf} package
    \citep{grf}.
  \item X-learner \citep{kunzel2019metalearners} uses the bootstrap to
    estimate the variance of CATE estimators. As with Causal Forest,
    the authors did not develop tools to cover ITE. The method is
    implemented in the \texttt{causalToolbox} package
    \citep{xlearner}.
  \item Bayesian Additive Regression Trees (BART) were initially
    developed as a flexible general-purpose Bayesian machine learning
    algorithm \citep{chipman2010bart}. They were later applied to
    causal inference \citep[e.g.][]{hill2011bayesian,
      green2012modeling, hahn2020bayesian} and found to outperform
    other methods both in terms of accuracy and coverage
    \citep{dorie2017aciccomp2016, dorie2019automated}. The method
    constructs Bayesian credible intervals to cover CATE.  By
    replacing credible intervals with prediction intervals, the method
    can be adapted to cover ITE. We use the functions
    \texttt{calc\_credible\_intervals} and
    \texttt{calc\_prediction\_intervals} from the \texttt{bartMachine}
    package \citep{bartMachine} to compute both intervals.
  \end{itemize}
  For weighted split-CQR, we estimate the propensity score via the
  gradient boosting algorithm \citep{friedman2001greedy} by using the
  \texttt{gbm} package \citep{gbm}. We further estimate the
  conditional quantiles in three different ways: (1) via quantile
  random forest \citep{athey2019generalized} by using the \texttt{grf}
  package \citep{grf}, (2) via quantile gradient boosting by using the
  \texttt{gbm} package \citep{gbm}, and (3) via the prediction
  posterior quantiles from BART by using the \texttt{bartMachine}
  package \citep{bartMachine}. For all conditional quantile
  estimators, we set
  $\alpha_{\lo} = \alpha / 2, \alpha_{\hi} = 1 - \alpha / 2$. Lastly,
  we use $75\%$ data as the training fold, as suggested by
  \cite{sesia2019comparison}. Our method is implemented in \texttt{R}
  \texttt{cfcausal} package, available at
  \url{https://github.com/lihualei71/cfcausal}. Code to replicate all
  the results from the paper is available at
  \url{https://github.com/lihualei71/cfcausalPaper}. 

\newcommand{\ntest}{n_{\text{test}}}

\begin{table}
  \caption{\label{tab:coverage}Summary of coverage guarantees in theory (left) and in our simulation study (right). }
  \centering
  \begin{tabular}{rllll}
    \toprule
    & CF & X-learner & BART & CQR\\
    \midrule
    CATE & \cmark & \cmark & \cmark & \xmark\\
    ITE & \xmark & \xmark & \cmark & \cmark\\
    \bottomrule
  \end{tabular}
  \hspace{1cm}
  \begin{tabular}{rllll}
    \toprule
    & CF & X-learner & BART & CQR\\
    \midrule
    CATE & \xmark & \xmark & \xmark & \cmark\\
    ITE & \xmark & \xmark & \xmark & \cmark\\
    \bottomrule
  \end{tabular}
\end{table}
    
  Each time, we generate $100$ independent datasets with sample size
  $n = 1000$. In each run, we also generate $\ntest = 10000$ extra
  independent data points, and construct 95\% confidence or prediction intervals for
  each of them via the aforementioned methods. We then estimate
  the empirical marginal coverage of CATE and ITE as
  $(1/\ntest)\sum_{i=1}^{\ntest}I(\tau(X_{i})\in \CITE(X_{i}))$ and
  $(1/\ntest)\sum_{i=1}^{\ntest}I(Y_{i}(1) - Y_{i}(0)\in
  \CITE(X_{i}))$,
  respectively, where $\CITE(X_{i}) = \hat{C}_{1}(X_{i})$ in this case.  Note that our metric is not asking for coverage
    at every point $x$. Instead, it demands coverage in an average
    sense. Therefore, a reliable method should, at the very least,
    have coverage close to or above $0.95$. To be sure, a method with
    invalid marginal coverage certainly cannot have valid conditional
    coverage. As summarized in Table \ref{tab:coverage} (left), note,
  and this is important, that Causal Forest and X-learner are only
  guaranteed to cover CATE, whereas weighted split-CQR is only
  guaranteed to cover ITE. Lastly, BART has guarantees to cover both
  CATE and ITE by employing two types of intervals.

\begin{figure}[H]
  \centering
  \includegraphics[width = 0.7\textwidth]{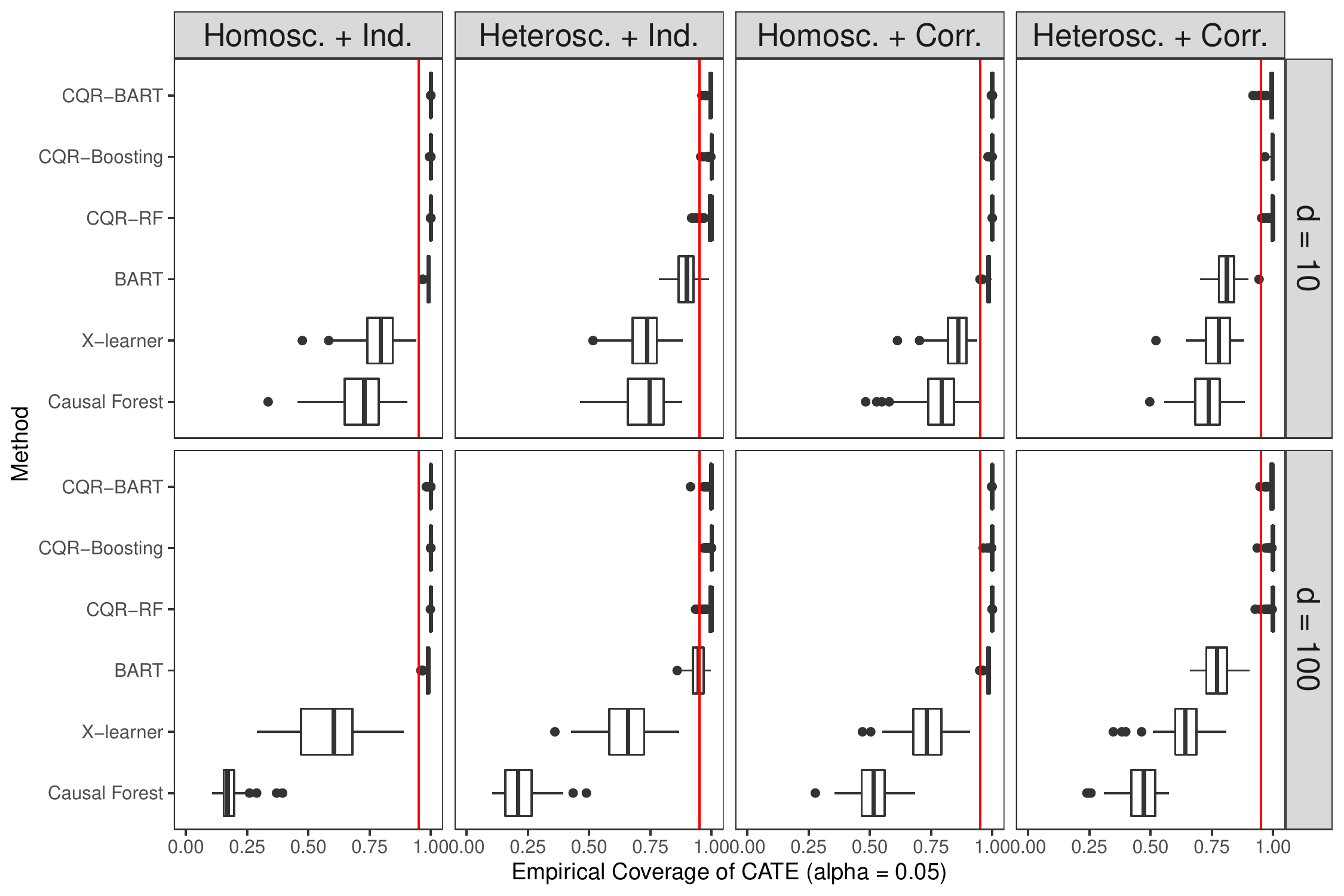}
  \caption{Empirical 95\% coverage of CATE. Each panel corresponds to
    one of the eight scenarios. ``Homosc.'' and ``Heterosc.'' are short for
    ``homoscedastic'' and ``heteroscedastic errors''; ``Ind.'' and
    ``Corr.'' are short for ``independent'' and ``correlated
    covariates''. }
\label{fig:simul1_coverage_tau}
\end{figure}

Figure \ref{fig:simul1_coverage_tau} presents CATE coverage results
for this simple example; recall that the model smoothly depends only
upon two variables out of ten or a hundred.  Causal Forest and
X-learner have poor coverage in all scenarios. Their performance
degrades even further in the higher dimensional setting $d =
100$.
Clearly, we must be far from the asymptotic setting considered in the
literature. BART has better coverage than Causal Forest and
X-learner. In the first three columns, we can see that the BART
credible intervals cover CATE. In the last column where the covariates
are correlated and the errors are heteroscedastic, BART has poor
coverage, especially in high dimensions. Although our method is not
guaranteed to cover CATE, we see that it achieves coverage in all
scenarios, although it may be conservative. Intuitively, this happens
because weighted split-CQR is designed to cover ITE and that in
reasonable models, prediction intervals are wider and often include
confidence intervals for the mean. In sum, CQR is the only method
achieving valid coverage in all scenarios (contrast this with the
theoretical predictions from Table \ref{tab:coverage} (left)).
  
  \begin{figure}[H]
  \centering
  \includegraphics[width = 0.7\textwidth]{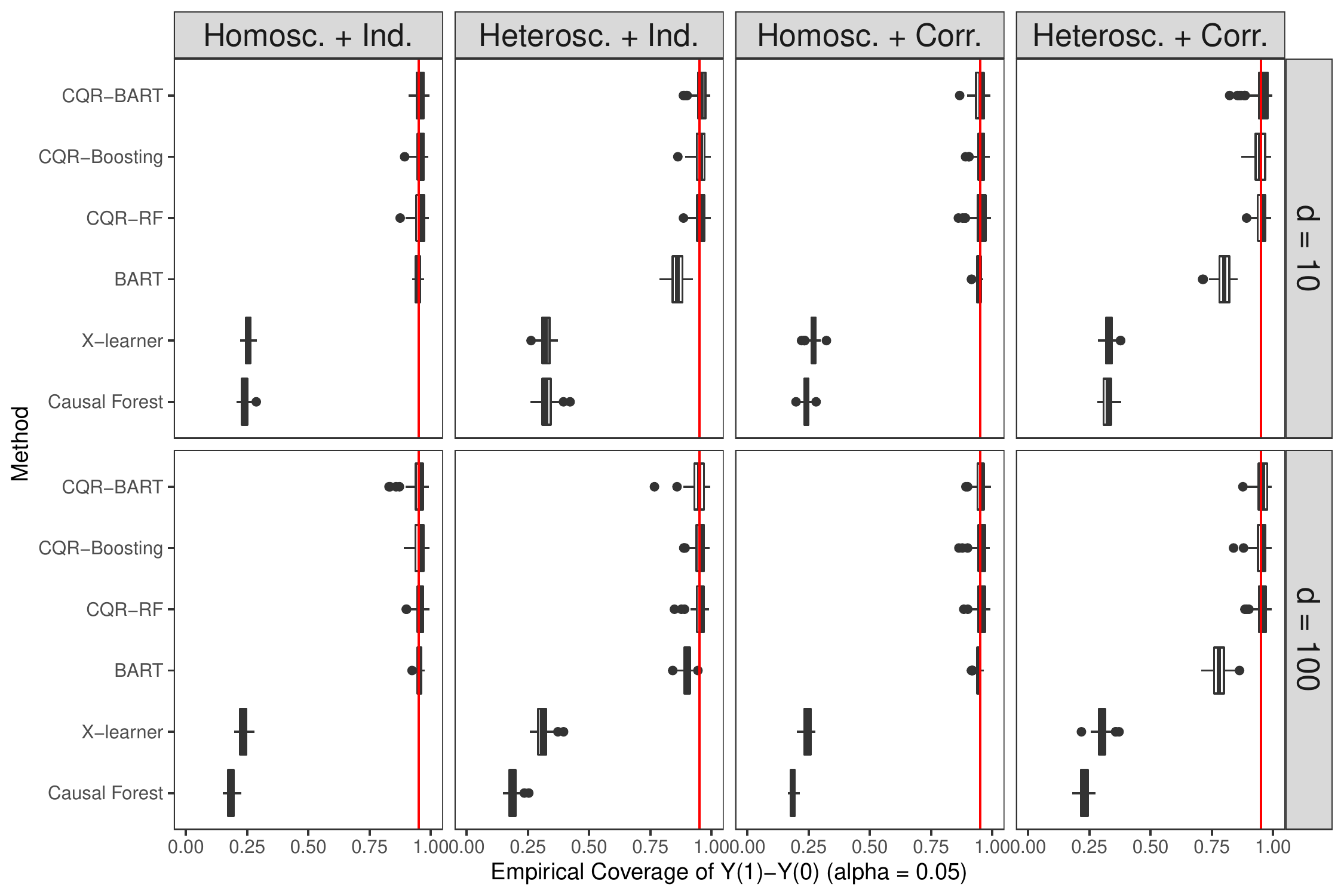}
  \caption{Empirical 95\% coverage of ITE. Everything else is as in
    Figure \ref{fig:simul1_coverage_tau}.} 
\label{fig:simul1_coverage}
\end{figure}

Figure \ref{fig:simul1_coverage} presents ITE coverage, which is
  the real subject of this paper. Causal Forest and X-learner  should not be regarded as competing methods since they are not
designed to cover ITE and it is therefore unsurprising that each
method has low coverage. We include the results here just to highlight the potential danger of misinterpreting the confidence intervals for CATE as ITE prediction intervals. The prediction intervals of BART have
perfect coverage with homoscedastic errors in both low and high
dimensions. However, in heteroscedastic cases, BART has unsatisfactory
coverage especially when the covariates are correlated. Finally, our
method achieves almost exact coverage regardless of the learning
procedures, regardless of whether the variables are correlated or not,
the `noise' is homoscedastic or not, and the dimension is low or high.

\begin{figure}[H]
  \centering
  \includegraphics[width = 0.7\textwidth]{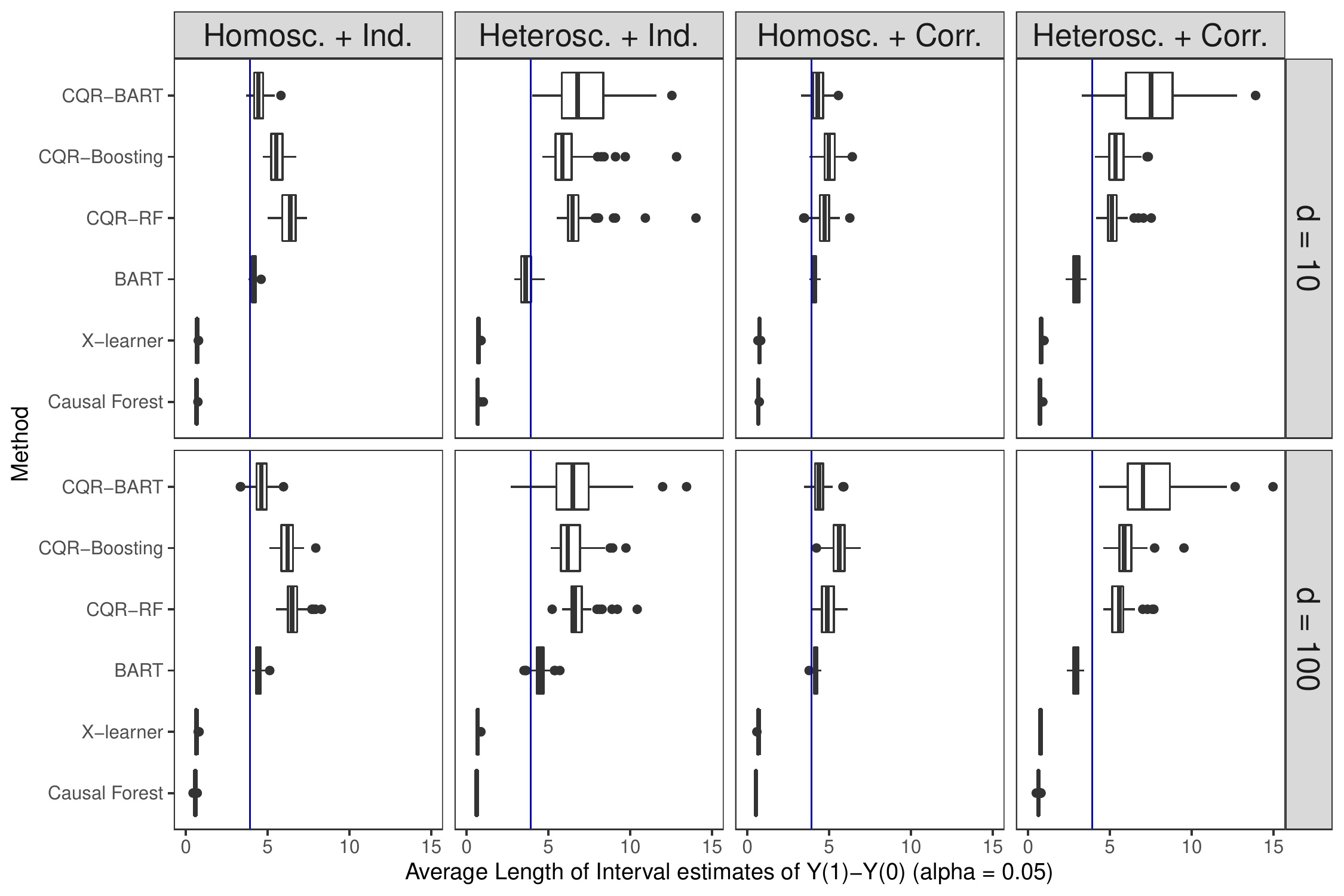}
  \caption{Lengths of interval estimates for ITE. The blue vertical line corresponds to the average length of oracle intervals. Everything else is as in
    Figure \ref{fig:simul1_coverage_tau}. } 
\label{fig:simul1_len}
\end{figure}

Next, we present interval lengths in Figure
\ref{fig:simul1_len}. Causal Forest and X-learner have short intervals
and we have seen that this is because they are poorly calibrated. In
homoscedastic settings where BART has valid coverage, BART also has
the shortest intervals. 
Notably, weighted split-CQR, with BART as
the learner of conditional quantiles, produces intervals that are
almost as narrow as those produced with BART. As explained in Section
\ref{subsec:double_robustness}, the correction $\eta(x) \approx 0$ since
BART fits the quantiles very well. As a result,
$\hat{C}(x) \approx [\hat{q}_{\alpha_{\lo}}(x),
\hat{q}_{\alpha_{\hi}}(x)]$.
We would like to observe that a major source of power loss is the data
splitting step since weighted split-CQR only uses 75\% data to train
BART. 
In heteroscedastic settings, BART has shorter intervals
because it has poor coverage as shown in Figure
\ref{fig:simul1_coverage}. We also see that weighted split-CQR has
much larger variability in interval lengths when using BART as the
learner. This is due to the fact that BART fails to estimate the
conditional quantiles well and thus yields a noisy conformity
correction $\eta(x)$.

To evaluate the tightness of these intervals, we compute the average
length of oracle intervals formed by the true $0.025$-th and
$0.975$-th conditional quantiles. In this case, the errors are
normally distributed and thus the expected length is
$(2 \times 1.96)\E [\sigma(X)]$. This is equal to $3.92$ in both cases
because $\int_{0}^{1}1dz = \int_{0}^{1}(-\log z)dz = 1$. In all
cases, we observe that the interval lengths of weighted split-CQR with
gradient boosting and random forest are reasonably short. In
homoscedastic cases, weighted split-CQR with BART almost achieves the
oracle length, despite having an expected sample size for inferring
$Y(1)$ equal to $n \E[e(X)] = (5/12)n \approx 417$.

\begin{figure}[H]
  \centering
  \includegraphics[width = 0.9\textwidth]{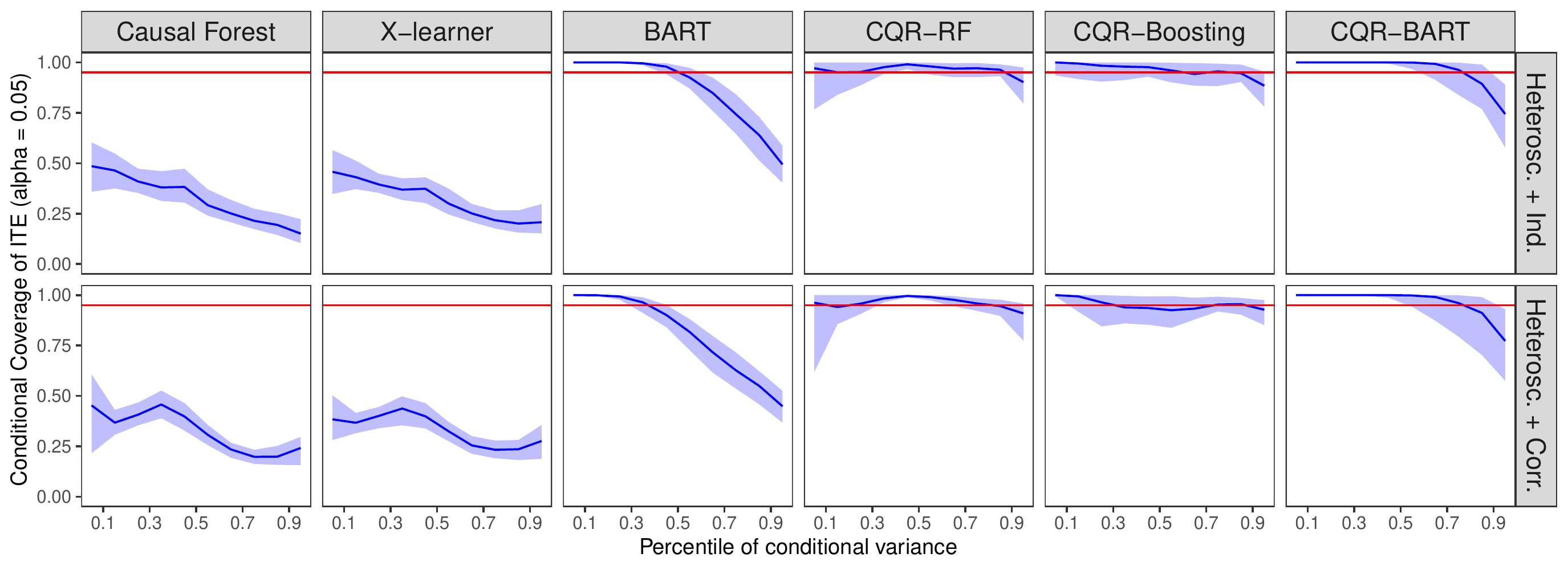}
  \caption{Estimated conditional coverage of ITE as a function of the
    conditional variance. Here, the dimension $d = 10$ and the
    coverage is set to 95\%. The blue curves correspond to the median
    and the blue confidence bands are the $5\%$ and $95\%$ quantiles
    of these estimates across $100$
    replicates.}\label{fig:synthetic_cond_tau}
\end{figure}

Finally, we turn to investigating the conditional coverage of all
these methods. 
Recall that in the heteroscedastic setting,
$\sigma^{2}(x) \rightarrow \infty$ as $x_{1}\rightarrow 0$.
Therefore, we expect the conditional coverage for instances with
larger values of $x_{1}$ to be lower. Figure
\ref{fig:synthetic_cond_tau} displays the estimated conditional
coverage of ITE as a function of the percentiles of $\sigma^{2}(x)$
when $d = 10$. Specifically, we stratify $\sigma^{2}(X_{i})$'s on the
$10000$ testing points into $10$ folds based on their
$10\%, 20\%, \ldots, 90\%$ percentiles and estimate the coverage
within each interval. It is clear that Causal Forest, X-learner and
BART all have decreasing conditional coverage as $\sigma^{2}(x)$
increases. Although BART has much better marginal coverage than the
other two methods, the poor conditional coverage near the right end
point is worrisome. In contrast, weighted split-CQR with quantile
random forest or quantile gradient boosting maintains conditional
coverage. While the weighted split-CQR with BART does not perform as
well as the other two variants, it improves upon BART implying that
the calibration is also helpful in securing conditional coverage. In
Appendix \ref{app:expr}, we show the same plots for $d = 100$ as well
as plots of conditional coverage stratified by the CATE function
$\tau(\cdot)$. They all exhibit similar patterns as shown in Figure
\ref{fig:synthetic_cond_tau}.

To clear any misinterpretation, we emphasize that our
  comparisons are based on interval estimates of ITE and CATE, and
  thus do not carry over to point estimates of CATE, which are of
  common interest in the literature. In general, the accuracy of a
  point estimate does not inform coverage of an interval
  estimate. Also, an interval estimate does not have to be derived
  from a point estimate and a weighted split-CQR conformal interval is
  an example.


\section{From Counterfactuals To Treatment Effects}\label{sec:effect}

We finally turn our attention to intervals for ITE for subjects not in
the study, and for which both potential oucomes are missing.

\subsection{A naive approach}\label{subsec:naive}

Consider an arbitrary testing point $x$.  Having constructed tools for
counterfactual inference producing interval estimates of a given
potential outcome, we can construct a pair of prediction intervals at
level $1-\alpha/2$, namely, $[\hat{Y}^{L}(1; x), \hat{Y}^{R}(1; x)]$
for $Y(1)$ and $[\hat{Y}^{L}(0; x), \hat{Y}^{R}(0; x)]$ for $Y(0)$. By
contrasting these two intervals, we can obtain an interval for ITE as
follows: 
\[\CITE(x) = [\hat{Y}^{L}(1; x) - \hat{Y}^{R}(0; x), \hat{Y}^{R}(1; x) - \hat{Y}^{L}(0; x)].\]
If the counterfactual intervals have guaranteed coverage, then
$\CITE(x)$ also covers ITE at the level $1-\alpha$ in the sense of
\eqref{eq:covITE}. 
This means that we can use
counterfactual intervals produced by weighted split-CQR or by BART
whenever they are suitable.

\subsection{A nested approach}\label{subsec:nest}
In this subsection,
we propose another strategy which we refer to as the nested approach. The nested procedure
starts by splitting the data into two folds. On the first fold we
train $\hat{C}_{1}(x)$ and $\hat{C}_{0}(x)$ by applying counterfactual
inference. On the second fold, for each unit $i$, we compute
$\hat{C}_{0}(X_{i})$ if $T_{i} = 1$ and compute $\hat{C}_{1}(X_{i})$
if $T_{i} = 0$. This induces an interval $\CITE(x; t, y^{\obs})$ for
ITE defined as
\[
\CITE(x; t, y^{\obs}) = 
    \begin{cases}
      y^{\obs} - \hat{C}_{0}(x), & t = 1,\\
      \hat{C}_{1}(x) - y^{\obs}, & t = 0. 
    \end{cases}
 \]
 Put $\hat{C}_{i} = \CITE(X_{i}; T_{i}, Y_{i}^{\obs})$  as in Table
 \ref{tab:twofolds}, which illustrates the procedure. For any unit $i$
 in the second fold,
\[
 \P(Y_{i}(1) - Y_{i}(0)\in \hat{C}_{i}) 
 = \P(T_{i} = 1)\P(Y_{i}(0)\in \hat{C}_{0}(X_{i}) \mid T_{i} = 1) + \P(T_{i} = 0)\P(Y_{i}(1)\in \hat{C}_{1}(X_{i}) \mid T_{i} = 0).
\]
This gives that if
\begin{equation}
  \label{eq:condition}
  \P(Y_{i}(0)\in \hat{C}_{0}(X_{i}) \mid T_{i} = 1) \ge 1 - \alpha, \quad \P(Y_{i}(1)\in \hat{C}_{1}(X_{i}) \mid T_{i} = 0)\ge 1 - \alpha,
\end{equation}
then
\begin{equation}
  \label{eq:fold2_cond}
  \P(Y_{i}(1) - Y_{i}(0)\in \hat{C}_{i})\ge 1 - \alpha.
\end{equation}

\begin{table}
  \caption{\label{tab:twofolds} Sketch of the two folds in the nested approach: the
    first fold (left) is used to construct counterfactual intervals;
    the second fold (right) is used to infer ITE.}  

  \scriptsize
  \begin{tabular}{ccc}
    \toprule
    $Y_{i}(1)$ & $Y_{i}(0)$ & $Y_{i}^{\mathrm{obs}}$\\
    \midrule
    \multicolumn{3}{l}{Treatment Group (fold 1)}\\
    \cmark & \xmark & $Y_{i}(1)$\\
    \midrule
    \multicolumn{3}{l}{Control Group (fold 1)}\\
    \xmark & \cmark & $Y_{i}(0)$\\
    \bottomrule
  \end{tabular}
  \hspace{0.5cm}
  \scriptsize  
  \begin{tabular}{ccc}
    \toprule
    $\hat{Y}_{i}(1)$ & $\hat{Y}_{i}(0)$ & $\hat{C}_{i}$ \\
    \midrule
    \multicolumn{3}{l}{Treatment Group (fold 2)}\\
    $Y_{i}(1)$ & $[\hat{Y}^{L}_{i}(0), \hat{Y}^{R}_{i}(0)]$ & $[Y_{i}(1) - \hat{Y}^{R}_{i}(0), Y_{i}(1) - \hat{Y}^{L}_{i}(0)]$\\
    \midrule
    \multicolumn{3}{l}{Control Group (fold 2)}\\
    $[\hat{Y}^{L}_{i}(1), \hat{Y}^{R}_{i}(1)]$ & $Y_{i}(0)$ & $[\hat{Y}^{L}_{i}(1) - Y_{i}(0), \hat{Y}^{R}_{i}(1) - Y_{i}(0)]$\\
    \bottomrule
  \end{tabular}
\end{table}

For randomized experiments with known propensity score $e(x)$,
\eqref{eq:condition} is satisfied if we use
$w_{0}(x) = e(x) / (1 - e(x))$ for $\hat{C}_{0}(x)$ and use
$w_{1}(x) = (1 - e(x)) / e(x)$ for $\hat{C}_{1}(x)$. Please note that
we do not need to split $\alpha$ for $\hat{C}_{0}(x)$ and
$\hat{C}_{1}(x)$. For observational studies, we can substitute $e(x)$
with $\hat{e}(x)$. By Theorem \ref{thm:double_robustness_ATE},
\eqref{eq:condition} holds approximately if either the propensity
scores or the conditional quantiles are estimated well.


The nested procedure creates an i.i.d.~dataset $(X_{i}, \hat{C}_{i})$,
conditional on the first fold, such that
$\hat{C}_{i} = \CITE(X_{i}; T_{i}, Y_{i}^{\obs})$ covers the ITE
with probability at least $1 - \alpha$. Therefore, the intervals
$\hat{C}_{i}$ serve as ``surrogate intervals'' for ITE. If we can fit
a model of $\hat{C}_{i}$ on $X_{i}$, denoted by $\tdCITE(x)$, then the
intervals can be generalized to subjects with both potential outcomes
missing. If $\tdCITE(X_{i})$ does not shrink $\hat{C}_{i}$ drastically,
it is likely that $\P(Y(1) - Y(0)\in \tdCITE(X_{i}))$ shall be close to
or above $1 - \alpha$.

It may be useful to think of the nested method as follows:
  instead of estimating the unobserved uncertainty, the nested method
  is fitting an observed ``uncertainty measurement''
  $\hat{C}_{i}$. This is arguably simpler.

\subsection{An inexact and an exact method under the nested framework}
The function $\tdCITE(x)$ can be obtained by training a model of the
left- and right-end point of $\hat{C}_{i}$ on $X_{i}$ separately using
generic machine learning methods. For instance, we can fit the 40\% quantile of the left-end point and 60\% quantile of the right-end point. Note that $\tdCITE$ does not have the
same theoretical guarantee as those offered by conformalized
counterfactual inference since the fit may not be controlled. For this
reason, we refer to it as an ``inexact method''. Nonetheless,  as will
be shown later, the inexact method achieves, in our empirical
examples, the target coverage with drastically shorter intervals than
the naive approach.

In extremely sensitive settings where the validity of predictions is
of serious concern, we may still need a method with a theoretical
guarantee of coverage. Here, we propose a secondary conformal
inference procedure on the induced dataset $(X_{i}, \hat{C}_{i})$.
Given a generic observation $(X, T, Y^{\obs})$, let
$C = \CITE(X, T, Y^{\obs})$ denote the induced interval. The second
procedure---the ``exact method''---is based on the simple observation
that if one can find an interval expansion function $\hat{\cC}(\cdot)$
that maps a covariate value to an interval such that
\begin{equation}
  \label{eq:interval_conformal_goal}
  \P(C\subset \hat{\cC}_\text{ITE}(X))\ge 1 - \gamma,
\end{equation}
then by \eqref{eq:fold2_cond}, 
\[\P(Y(1) - Y(0) \not\in \hat{\cC}_\text{ITE}(X))\le \P(Y(1) - Y(0) \not\in C) + \P(C\not\subset \hat{\cC}_\text{ITE}(X))\le \alpha + \gamma.\]

\begin{algorithm}[H]
  \caption{(Unweighted) conformal inference for interval outcomes}  \label{algo:int}
  \textbf{Input: } level $\gamma$, data $\Z = (X_{i}, C_{i})_{i\in \I}$ where $C_{i} = [C_{i}^{L}, C_{i}^{R}]$, testing point $x$, 

  \hspace{1.3cm} functions $\hat{m}^{L}(x; \D), \hat{m}^{R}(x; \D)$ to fit the conditional mean/median of $C^{L}, C^{R}$

  \textbf{Procedure: }
  \begin{algorithmic}[1]
    \State Split $\Z$ into a training fold $\Z_{\tr} \triangleq (X_{i}, C_{i})_{i\in \I_{\tr}}$ and a calibration fold $\Z_{\ca}\triangleq (X_{i}, C_{i})_{i\in \I_{\ca}}$
    \State For each $i \in \I_{\ca}$, compute score $V_{i} = \max\{\hat{m}^{L}(X_{i}; \Z_{\tr}) - C_{i}^{L}, C_{i}^{R} - \hat{m}^{R}(X_{i}; \Z_{\tr})\}$
    \State Compute $\eta$ as the $(1 - \gamma)(1 + 1 / |\Z_{\ca}|)$ quantile of the empirical distribution of $\{V_{i}: i\in \I_{\ca}\}$
  \end{algorithmic}

  \textbf{Output: $\hat{\cC}(x) = [\hat{m}^{L}(x; \Z_{\tr}) - \eta, \hat{m}^{R}(x; \Z_{\tr}) + \eta]$}
\end{algorithm}

Denote by $C^{L}$ and $C^{R}$ the left- and the right-end point of
$C$, respectively. To achieve \eqref{eq:interval_conformal_goal}, we
need to find a lower prediction bound for $C^{L}$ and an upper
prediction bound for $C^{R}$. A naive approach is to apply standard
(unweighted) one-sided conformal inference on $C^{L}$ and $C^{R}$ with
level $\gamma / 2$ separately. Such a crude Bonferroni correction may
be conservative in practice. To overcome this, we propose a conformal
inference procedure that jointly calibrates $C^{L}$ and $C^{R}$. The
coverage guarantee can be proved using the standard argument
\citep[e.g.][]{lei2018distribution, romano2019conformalized} and we
include the proof in Appendix \ref{app:miscellaneous} for
completeness. 
\begin{theorem}\label{thm:int_conformal}
  Consider Algorithm \ref{algo:int} and assume $(X_{i}, C_{i})\stackrel{i.i.d.}{\sim} (X, C)$. Then
  \[\P(C\subset \hat{\cC}(X))\ge 1 - \gamma.\]
\end{theorem}

With all this in place, both the inexact and the exact methods are stated in
Algorithm \ref{algo:two_step}.
\begin{algorithm}
  \caption{Nested approach for interval estimates of ITE}  \label{algo:two_step}
  \textbf{Input: } level $\alpha$, level $\gamma$ (only for exact version), data $\Z = (X_{i}, Y_{i}, T_{i})_{i=1}^{n}$, testing point $x$

  \vspace{0.2cm}
  \textbf{Step I: data splitting}
  \begin{algorithmic}[1]
    \State Split the data into two folds $\Z_{1}$ and $\Z_{2}$
    \State Estimate propensity score $\hat{e}(x)$ on $\Z_{1}$
  \end{algorithmic}

  \vspace{0.2cm}
  \textbf{Step II: counterfactual inference on $\Z_{2}$}
  \begin{algorithmic}[1]   
    \For{$i$ in $\Z_{2}$ with $T_{i} = 1$}
         \State Compute $[\hat{Y}_{i}^{L}(0), \hat{Y}_{i}^{R}(0)]$ in Algorithm \ref{algo:split-CQR} on $\Z_{1}$ with level $\alpha$ and $w_{0}(x) = \hat{e}(x) / (1 - \hat{e}(x))$
         \State Compute $\hat{C}_{i} = [Y_{i}(1) - \hat{Y}_{i}^{R}(0), Y_{i}(1) - \hat{Y}_{i}^{L}(0)]$
    \EndFor
    \For{$i$ in $\Z_{2}$ with $T_{i} = 0$}
         \State Compute $[\hat{Y}_{i}^{L}(1), \hat{Y}_{i}^{R}(1)]$ in Algorithm \ref{algo:split-CQR} on $\Z_{1}$ with level $\alpha$ and $w_{1}(x) = (1 - \hat{e}(x)) / \hat{e}(x)$
         \State Compute $\hat{C}_{i} = [\hat{Y}_{i}^{L}(1) - Y_{i}(0), \hat{Y}_{i}^{R}(1) - Y_{i}(0)]$
    \EndFor
  \end{algorithmic}

  \vspace{0.2cm}
  \textbf{Step III: Interval of ITE on the testing point}
  \begin{algorithmic}[1]
    \State (Exact version) Apply Algorithm \ref{algo:int} on $(X_{i}, \hat{C}_{i})_{i\in \Z_{2}}$ with level $\gamma$, yielding an interval $\hat{\cC}_\text{ITE}(x)$
    \State (Inexact version) Fit conditional quantiles of $\hat{C}^{L}$ and $\hat{C}^{R}$, yielding an interval $\hat{\cC}_\text{ITE}(x)$
  \end{algorithmic}
  \textbf{Output: $\hat{\cC}_\text{ITE}(x)$}\label{algo:nest_CQR}
\end{algorithm}

\subsection{Empirical performance}

To evaluate the performance of our methods, we design numerical
experiments on the data analyzed in the 2018 Atlantic Causal Inference
Conference workshop on heterogeneous treatment effects
\citep{carvalho2019assessing}. The workshop organizers generated a
synthetic dataset based on the National Study of Learning Mindsets
(NLSM) \citep{yeager2019national}, a large-scale randomized trial of a
behavioral intervention, to emulate an observational study.  For
information on the dataset, please see Section 2 from
\cite{carvalho2019assessing}. Due to privacy concerns, the workshop
organizers only released limited information on the data generating
process, as well as the simulated dataset, available at
\url{https://github.com/grf-labs/grf/tree/master/experiments/acic18}.
Although the focus of this workshop was
on heterogeneous treatment effects, the organizers did not evaluate
whether the submissions cover the ITE or CATE. In this subsection, we
shall fill in this gap by comparing our method with Causal Forest,
X-learner and BART, just as we did in Section \ref{subsec:simul}.

Obviously, we must know the ground truth in order to evaluate
coverage. Therefore, we generated synthetic datasets based on the
available information from \cite{carvalho2019assessing}. First, we
split data into two folds $\Z_{1}$ and $\Z_{2}$, with
$|\Z_{1}| = 2079$ including 20\% of the samples and $|\Z_{2}| = 8312$
including the remaining 80\%. In our numerical experiments, we
generate the covariate vector $X$ by sampling from $\Z_{2}$ with
replacement. To generate the potential outcomes, we apply random
forest from \texttt{R} \texttt{randomForest} package on $\Z_{1}$ to
fit $\E [Y(0)]$. Denote the output by $\hat{m}_{0}(x)$. Then we
generate $\E [Y(1)]$ by adding the CATE function $\tau(x)$ (equation
(1) of \cite{carvalho2019assessing}) to $\hat{m}_{0}(x)$. To account
for heteroscedasticity, we apply quantile random forest from
\texttt{R} \texttt{grf} package to fit the 25\% and 75\% conditional
quantiles of $Y(0)$ and $Y(1)$ and compute the conditional
interquartile ranges $\hat{r}_{0}(x)$ and $\hat{r}_{1}(x)$. Given a
covariate vector $X_{i}$, we subsequently generate $Y_{i}(1)$ and
$Y_{i}(0)$ as
\[
Y_{i}(1) = \hat{m}_{0}(X_{i}) + \tau(X_{i}) +
0.5\hat{r}_{1}(X_{i})\eps_{i1}, \quad Y_{i}(0) = \hat{m}_{0}(X_{i}) +
0.5\hat{r}_{0}(X_{i})\eps_{i0},\quad \eps_{i1},
\eps_{i0}\stackrel{i.i.d.}{\sim} N(0, 1).
\]
Finally, we generate propensity scores by applying random forest from
\texttt{R} \texttt{randomForest} package on $\Z_{1}$ and truncating
the estimated propensity score $\hat{e}(x)$ at $0.1$ and $0.9$ to
guarantee overlap. For each $X_{i}$, we generate $T_{i}$ as a
Bernoulli random variable with parameter $\hat{e}(X_{i})$.

For each run, we first generate $6000$ quadruples
$(X_{i}, T_{i}, Y_{i}(1), Y_{i}(0))$ sampled according to the above
data generating process. We then randomly select $n = 1000$ samples as
the training set to induce an observational study with observations
$(X_{i}, T_{i}, Y_{i}^{\obs})$. For the remaining $5000$ testing
samples, only the covariates $X_{i}$ are accessible to the analyst;
the triple $(T_{i}, Y_{i}(1), Y_{i}(0))$ is solely used to evaluate
coverage. It goes without saying that each method will take the
training set as the input and, for each testing sample, produce $95\%$
intervals for ITE.

For weighted split-CQR, we apply the naive procedure as described in Section \ref{subsec:naive} and the nested procedure with both exact and inexact calibration as described in Algorithm \ref{algo:nest_CQR}. For all procedures, we apply BART to fit conditional quantiles, as in Section \ref{subsec:simul}, and apply gradient boosting from \texttt{R} \texttt{gbm} package to fit propensity scores. For the exact nested method, we set $\alpha = \gamma = 0.025$. 

Note that neither the naive nor the nested approaches are limited to
CQR. Therefore, we also wrap both methods around BART as competitors,
where only the inexact version is considered for the nested approach
since BART cannot produce exact counterfactual intervals. Finally,
since Causal Forest and X-learner cannot produce counterfactual
intervals, except in the special case from Section \ref{subsec:simul}
where $Y(0)\equiv 0$, we do not wrap the naive or the nested methods
around them but directly report their confidence intervals as
benchmarks instead. 

\begin{figure}[H]
  \centering
  \includegraphics[width = 0.48\textwidth]{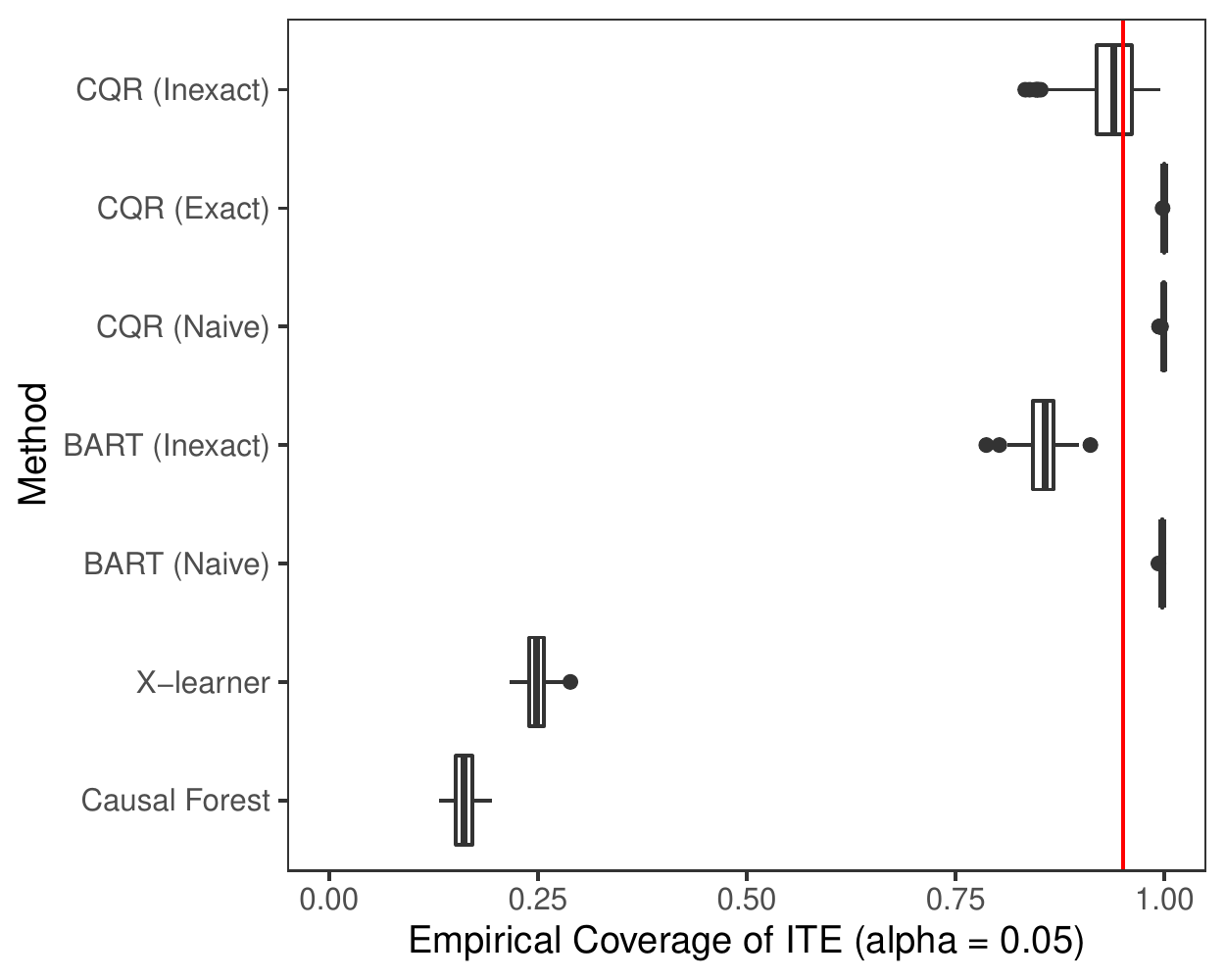}
  \includegraphics[width = 0.48\textwidth]{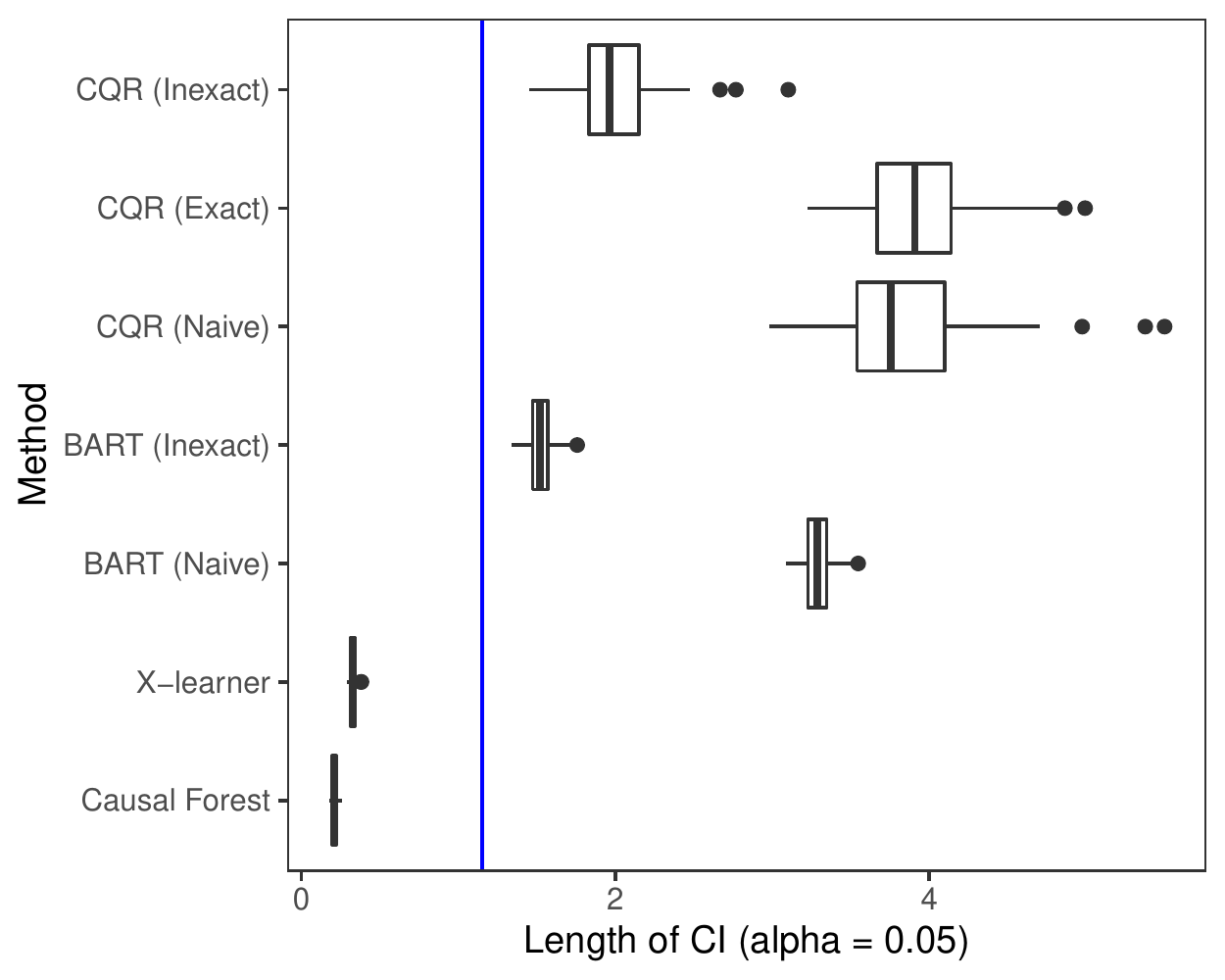}
  \caption{Coverage (left) and average length (right) of intervals for ITE on synthetic data generated from the NLSM data. The red vertical line corresponds to the target coverage $95\%$ and the blue vertical line corresponds to the average length of oracle intervals.}\label{fig:NLSM_simul_ITE}
\end{figure}

\begin{figure}[H]
  \centering
  \includegraphics[width = 0.9\textwidth]{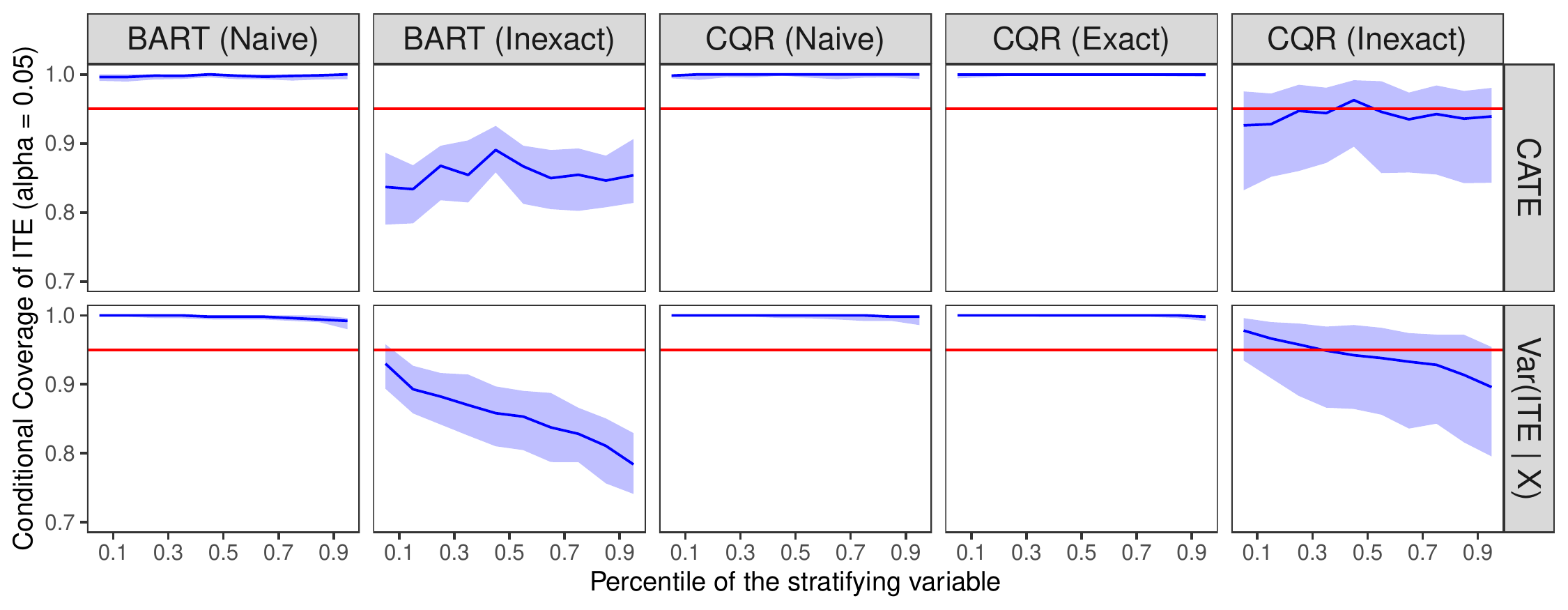}
  \caption{Estimated conditional coverage of ITE as a function of the
    conditional variance $\sigma^{2}(x)$ (upper row), and CATE
    $\tau(x)$ (lower row). The coverage is set to $95\%$. The blue
    curves correspond to the median and the blue confidence bands are
    the $5\%$ and $95\%$ quantiles of these estimates across $100$
    replicates.}\label{fig:NLSM_cond}

\end{figure}

Figure \ref{fig:NLSM_simul_ITE} presents the coverage and the average
length of intervals estimated on the testing set by repeating the
above procedure 100 times. As expected, the naive methods with both
CQR and BART are conservative. The exact nested method with CQR is
also conservative although the coverage is only guaranteed to be above
$95\%$ in the worst case. In contrast, the inexact nested methods with
CQR or BART are less conservative. Notably, BART fails to achieve the
desired coverage but CQR, with BART as the learner, calibrates it
successfully. Also, we can see from the right panel that the average
length of intervals of inexact-CQR is just slightly above that of
inexact-BART while significantly lower than that of either the naive
or the exact nested methods. Moreover, in accordance with the
observations from Section \ref{subsec:simul}, Causal Forest and
X-learner have poor coverage and, thus, their (short) intervals are
misleading. As in Section \ref{subsec:simul}, we compute the average
length of oracle intervals as $3.92\E [\sigma(X)]$ where
$\sigma(x) = 0.5\sqrt{\hat{r}_{1}(x)^{2} + \hat{r}_{0}(x)^{2}}$.
Having said this, we are in a different situation here since the true
conditional quantiles of ITE can never be identified without
assumptions on the joint distribution of $Y(1)$ and $Y(0)$.
Therefore, this oracle length cannot be achieved in general no matter
how powerful the model fitting. Keeping this cautionary remark in
mind, we nonetheless see that weighted split-CQR still produces
reasonably short intervals.

Finally, as in Section \ref{subsec:simul}, we investigate the
conditional coverage as a function of the conditional variance
$\sigma^{2}(x)$ and the CATE $\tau(x)$, respectively. For better
visualization, we exclude Causal Forest and X-learner since they have
poor marginal coverage. We can see from Figure \ref{fig:NLSM_cond}
that inexact-CQR has desirable and relatively even conditional
coverage while inexact-BART performs worse.

\subsection{Re-analysing NLSM data}
In this subsection, we apply inexact-CQR with BART as the learner to re-analyze the NLSM data from \cite{carvalho2019assessing}. Since the ground truth is inaccessible, we only perform an exploratory analysis for the purpose of illustration.

\begin{figure}
  \centering
  \includegraphics[width = 0.32\textwidth]{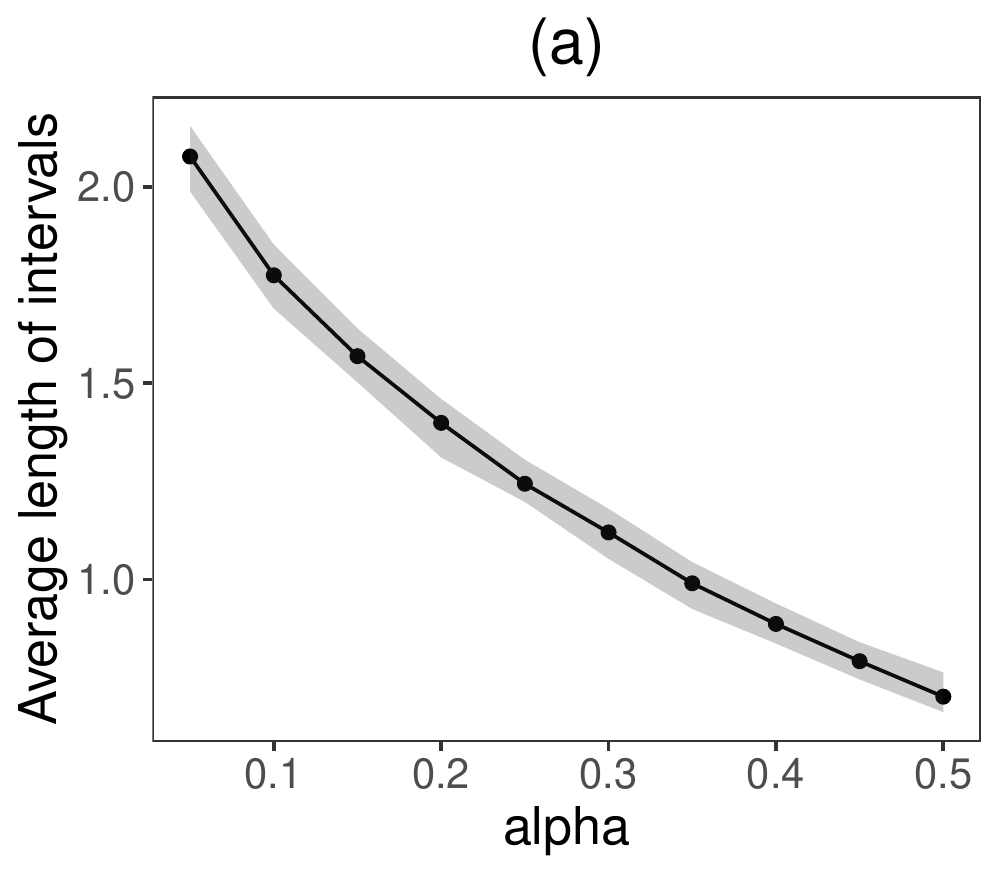}
  \includegraphics[width = 0.32\textwidth]{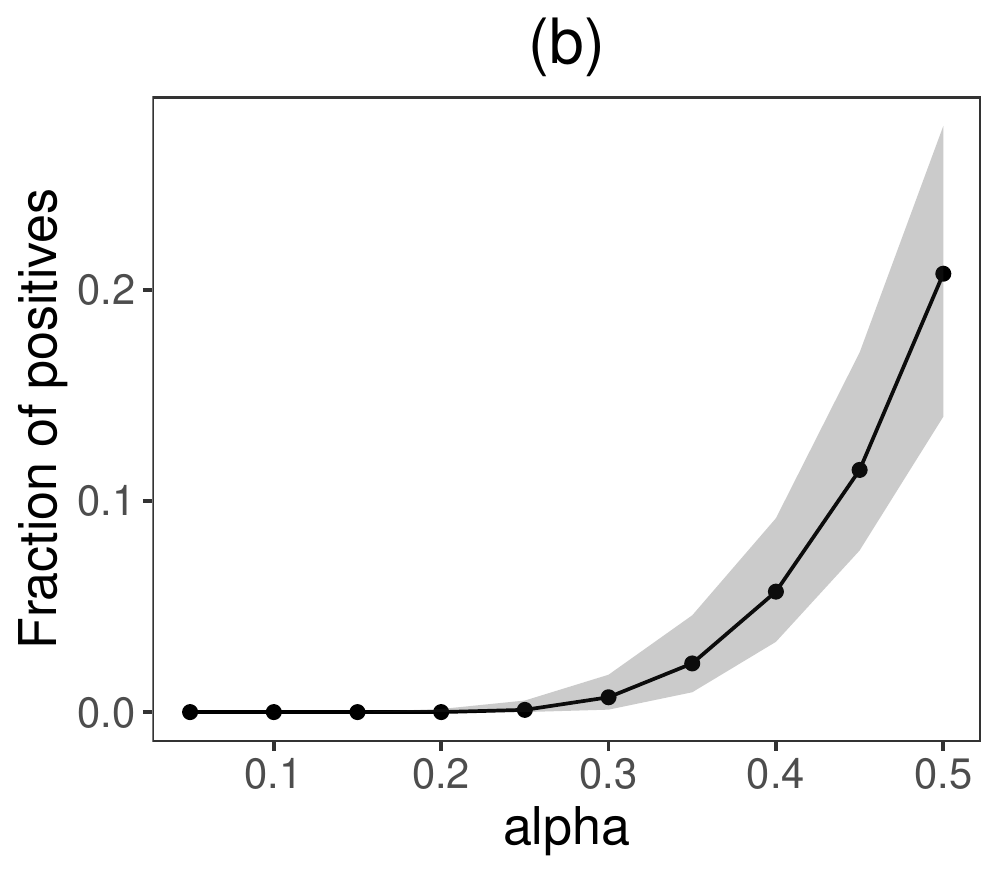}
  \includegraphics[width = 0.32\textwidth]{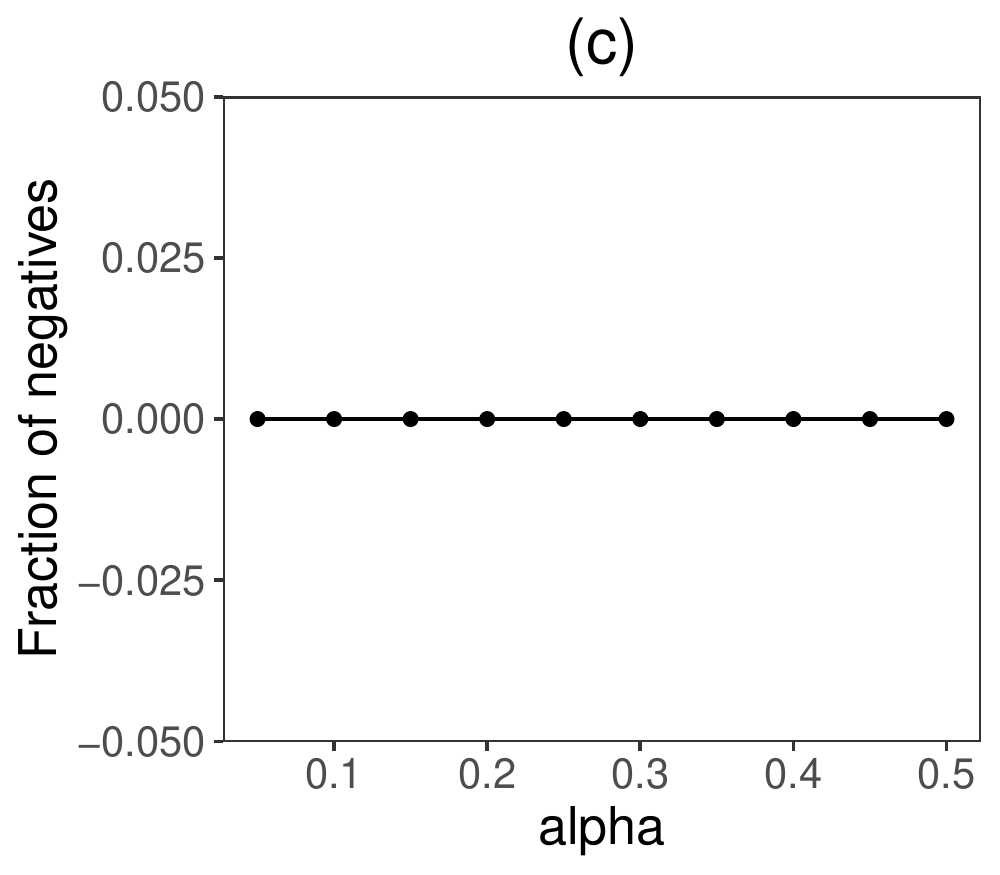}
  \caption{Re-analysis of NLSM data: (a) average length of intervals; (b) fraction of intervals with positive lower bounds; (c) fraction of intervals with negative upper bounds}\label{fig:NLSM_analysis}  
\end{figure}

To create informative testing points, we split the data into two folds
$\Z_{1}$ and $\Z_{2}$. Then we apply weighted inexact-CQR on $\Z_{1}$
to produce intervals for ITE for each point in $\Z_{2}$. Similarly, we
apply weighted inexact-CQR on $\Z_{2}$ to produce intervals of ITE for
each point in $\Z_{1}$. To account for the variability from data
splitting, we repeat the above procedure $100$ times. Figure
\ref{fig:NLSM_analysis} (a) displays the average length of intervals
as a function of level $\alpha$ with the upper and lower envelopes
being respectively the $95\%$ and $5\%$ quantiles across $100$ runs.

With these intervals, we can make inferential claims on ITE with
confidence. For instance, we can decide to assign treatment to a
patient if the lower prediction bound of her ITE interval is
positive. Since these intervals are guaranteed to have desired
coverage, it is likely that they produce fewer false positives on
  the average, although we leave a theoretical investigation of this
intuition to future work. Figure \ref{fig:NLSM_analysis} (b) and (c)
show the fractions of intervals that only cover positive and negative
values, respectively. We see some evidence of positive ITE when
$\alpha$ is above $0.25$ while no evidence of any negative ITE even
when $\alpha = 0.5$.


\section{From Potential Outcomes to Other Causal Frameworks}\label{sec:extension}
We proposed a method based on weighted conformal inference which
produces interval estimates of counterfactuals and individual
treatment effects under the potential outcome framework. For
randomized experiments with perfect compliance, our method has
guaranteed coverage in finite samples without any modeling assumptions
on the data generating process. For randomized experiments with
ignorable compliance or general observational studies under the strong
ignorability assumption, our method is doubly robust in the sense that
the coverage is asymptotically guaranteed if either the conditional
quantiles of potential outcomes or the propensity scores are
consistently estimated. In contrast, existing methods may suffer from
a significant coverage deficit even in simple models. Furthermore, our
framework naturally extends to other populations by modifying the
weight function according to Table \ref{tab:weight_func}.

The key observation is the covariate shift together with the
invariance of the conditional distribution: the observed distribution
of $(X, Y(1))$ is $P_{X\mid T = 1}\times P_{Y(1)\mid X}$ under
ignorability while the target distribution is
$P_{X}\times P_{Y(1)\mid X}$. Now the invariance of the conditional distribution is
also the enabling property in other frameworks of causal inference. As
a consequence, our method can be naturally extended to those settings,
and we close this paper by discussing two of them.

\subsection{Causal diagram framework}
Judea Pearl in his pioneering work \citep{pearl1995causal} introduced
a general framework of causal inference based on graphical models,
which has been widely applied, see
  e.g.~\cite{greenland1999causal, spirtes2000causation,
    richardson2013single, glymour2014discovering,
    tennant2019use}. This framework does not rely on the notion of
counterfactuals but instead defines causal effects through the
\texttt{do} operator, which modifies the observed distribution by
removing the causal paths that directly point to the intervention
variable. We refer the readers to \cite{pearl2018book} for the
philosophy and basics of causal diagrams and to \cite{pearl2016causal}
for the mathematical foundation of this framework.

We here only discuss the case where $T$ is the intervention variable,
$Y$ is the outcome variable and $X$ is a set of variables that
satisfies the back-door criterion. Roughly speaking, this means that $X$
includes all confounders and excludes all post-treatment variables. In
this case, the foundational result in \cite{pearl1995causal} shows
that
\[P_{(X, Y) \mid \mathrm{do}(T = t)} = P_{X}\times P_{Y\mid X, T = t}.\]
This is the target distribution to be inferred. In contrast, the observed distribution of $(X, Y)$ given $T = t$ is
\[P_{(X, Y)\mid T = t} = P_{X\mid T = t}\times P_{Y\mid X, T = t}.\]
Clearly, this has exactly the same structure as in the potential
outcome framework. As a consequence, weighted split-CQR can be applied
without any modification to produce doubly robust interval estimates
of $Y$ under the \texttt{do} intervention.

\subsection{Invariant prediction framework}

Invariant prediction is another framework proposed by
\cite{peters2016causal}. It is particularly powerful when there are
multiple data sources under different interventions, such as in gene
knockout experiments. Consider an outcome variable $Y$, a set of
interventions or covariates $X$ and an environment variable $E$ that
indicates the source of data. Then one basic setting under this
framework assumes that $Y\indep E \mid X$ while $X$ may depend on $E$ with
$X \mid E\sim P_{X}^{E}$. The goal is to predict the outcome under a
new environment. For simplicity, we assume that a dataset
$(X_{ij}, Y_{ij})_{i=1}^{n_{j}}$ is available for each environment
$e_{1}, \ldots, e_{J}$ and a testing dataset $(X_{i0})_{i=1}^{n_{0}}$
for the target environment $e_{0}$. Due to the invariance assumption,
on the $j$-th dataset, the observed distribution of $(X, Y)$ is
$P_{X}^{e_{j}}\times P_{Y\mid X}$ while the target distribution to be
inferred on is $P_{X}^{e_{0}}\times P_{Y\mid X}$.

Again, this is reduced to a problem of constructing valid prediction
intervals under covariate shifts. When $J = 1$, this has exactly the
same structure as in the potential outcome framework and thus weighted
split-CQR with weight function $dP_{X}^{e_{0}}(x) / dP_{X}^{e_{1}}(x)$
produces doubly robust intervals of $Y$ under environment $e_{0}$. When $J > 1$, we can in principle apply the general
weighted conformal inference techniques from
\cite{barber2019conformal}. However the weight function becomes much
more complicated than that in Algorithm \ref{algo:split-CQR}. An
alternative approach is to create a weighted population from
environments $e_{1}, \ldots, e_{J}$ with observed distribution
\[\lb\sum_{j=1}^{J}q_{j}P_{X}^{e_{j}}\rb\times P_{Y\mid X},\]
and apply Algorithm \ref{algo:split-CQR} on this pseudo dataset with weight $w(x) = 1 / \sum_{j=1}^{J}q_{j}(dP_{X}^{e_{j}} / dP_{X}^{e_{0}})(x)$. The weights $q_{j}$ can be chosen through certain balancing procedures that forces the covariate distribution to approximate $P_{X}^{e_{0}}$. We leave the formal development of this idea to future work.

\section*{Acknowledgment}
E. C. was supported by Office of Naval Research grant N00014-20-12157. L. L. was supported by NSF OAC grant 1934578. 
The authors are grateful to David Arbour, Susan Athey, Eytan Bakshy, Stephen Bates, Richard A. Berk, Peter J. Bickel, Jelena Bradic, Andreas Buja, Michael Celentano, Peng Ding, Dean Eckles, Nikolaos Ignatiadis, Yucen Luo, Edward Kennedy, Roger Koenker, Arun Kumar Kuchibhotla, Elizabeth L. Ogburn, Zhimei Ren, Thomas S. Richardson, James Robins, Chiara Sabatti, Jasjeet S. Sekhon, Dylan Small, Sean Taylor, Eric Tchetgen Tchetgen, Yuhao Wang, Jeffrey Wong, Bin Yu, and Xiao-Hua Zhou for their constructive feedback.

\bibliography{conformal_causal}
\bibliographystyle{plainnat}

\clearpage
\appendix

~\\
\begin{center}
  \begin{Large}
    \textbf{Appendix}
  \end{Large}
\end{center}

\section{Nonasymptotic Theory for Double Robustness of Weighted Split-CQR}\label{app:proofs}
This section establishes the double robustness of general weighted
split-CQR. We first prove nonasymptotic results for two sides of the double robustness and then present a simpler asymptotic result as a corollary in Section \ref{subsec:asymptotic}.

\begin{theorem}\label{thm:double_robustness_w}
  Let
  $(X_{i}, Y_{i})\stackrel{i.i.d.}{\sim}(X, Y) \sim P_{X}\times
  P_{Y\mid X}$ and $Q_{X}$ be another distribution on the domain of
  $X$. Set $N = |\Z_{\tr}|$ and $n = |\Z_{\ca}|$. Further, let
  $\hat{q}_{\beta, N}(x) = \hat{q}_{\beta, N}(x; \Z_{\tr})$ be an
  estimate of the $\beta$-th conditional quantile $q_{\beta}(x)$ of
  $Y \mid X = x$, $\hat{w}_{N}(x) = \hat{w}_{N}(x; \Z_{\tr})$ be an
  estimate of $w(x) = (dQ_{X} / dP_{X})(x)$, and $\hat{C}_{N, n}(x)$
  be the conformal interval resulting from Algorithm
  \ref{algo:split-CQR}. Assume that
  $\E[\hat{w}_{N}(X) \mid \Z_{\tr}] < \infty$, where $\E$ denotes
  expectation over $X\sim P_{X}$. Redefine
  $\hat{w}_{N}(x)$ as
  $\hat{w}_{N}(x) / \E[\hat{w}_{N}(X) \mid \Z_{\tr}]$ so that
  $\E[\hat{w}_{N}(X) \mid \Z_{\tr}] = 1$. Then
  \begin{equation}\label{eq:unconditional_coverage_rate_w}
    \P_{(X, Y)\sim Q_{X}\times P_{Y\mid X}}\lb Y\in \hat{C}_{N, n}(X)\rb\ge 1 - \alpha - \frac{1}{2}\E_{X\sim P_{X}}|\hat{w}_{N}(X) - w(X)|.
  \end{equation}

  \end{theorem}

  \begin{theorem}\label{thm:double_robustness_q}
  In the setting of Theorem \ref{thm:double_robustness_w}, assume further that
    \begin{enumerate}[(1)]
    \item $\alpha_{\hi} - \alpha_{\lo} = 1 - \alpha$;
    \item there exist $r, b_{1}, b_{2} > 0$ such that $\P(Y = y \mid X = x) \in [b_{1}, b_{2}]$ uniformly over all $(x, y)$ with $y\in [q_{\alpha_{\lo}}(x) - r, q_{\alpha_{\lo}}(x) + r]\cup [q_{\alpha_{\hi}}(x) - r, q_{\alpha_{\hi}}(x) + r]$;
    \item $\P_{X\sim Q_{X}}(w(X) < \infty) = 1$, and there exist $\delta, M > 0$ such that $\lb\E\left[\hat{w}_{N}(X)^{1 + \delta}\right]\rb^{1 / (1 + \delta)} \le M$;
    \item there exist $k, \ell > 0$ such that $\displaystyle \lim_{N\rightarrow \infty} \E[\hat{w}_{N}(X)H_{N}^{k}(X)] = \lim_{N\rightarrow \infty} \E[w(X)H_{N}^{\ell}(X)] = 0$, where
      \[\displaystyle H_{N}(x) = \max\{|\hat{q}_{\alpha_{\lo}, N}(x) - q_{\alpha_{\lo}}(x)|, |\hat{q}_{\alpha_{\hi}, N}(x) - q_{\alpha_{\hi}}(x)|\}.\]
    \end{enumerate}
    Then there is a constant $\const_{1}$ that only depends on $r, b_1, b_2, \delta, M, k, \ell$ such that
    \begin{multline}
      \P_{(X, Y)\sim Q_{X}\times P_{Y\mid X}}(Y \in \hat{C}_{N, n}(X))\ge 1 - \alpha \\
       -\const_{1} \left\{\frac{(\log n)^{(1 + \delta') / 2(2 + \delta')}}{n^{\delta' / (2 + \delta')}} + \lb\E[\hat{w}_{N}(X)H_{N}^{k}(X)]\rb^{1 / (2+k)} + \lb\E[w(X)H_{N}^{\ell}(X)]\rb^{1/ (1+\ell)}\right\},\label{eq:unconditional_coverage_rate_q}
    \end{multline}
    where $\delta' = \min\{\delta, 1\}$. Furthermore, for any $\beta\in (0, 1)$, there is a constant $\const_{2}$ that only depends on $r, b_1, b_2, \delta, M, k, \ell, \beta$ such that, with probability at least $1 - \beta$, 
    \begin{multline}
      \P_{(X, Y)\sim Q_{X}\times P_{Y\mid X}}(Y \in \hat{C}_{N, n}(X) \mid X )\ge 1 - \alpha \\
       - \const_{2}\left\{\frac{(\log n)^{(1 + \delta') / 2(2 + \delta')}}{n^{\delta' / (2 + \delta')}} + \lb\E[\hat{w}_{N}(X)H_{N}^{k}(X)]\rb^{1 / (2+k)} + \lb\E[w(X)H_{N}^{\ell}(X)]\rb^{1/ \ell}\right\}.\label{eq:conditional_coverage_rate_q}
    \end{multline}
  \end{theorem}



  \begin{remark}
    If $\delta \ge 1$ and $H_{N}(X)\le \Delta_{N}$ almost surely for some deterministic sequence $\Delta_{N} = o(1)$, by letting $k, \ell\rightarrow \infty$, the RHS of \eqref{eq:unconditional_coverage_rate_q} and \eqref{eq:conditional_coverage_rate_q} reduce to
    \[1 - \alpha - \const \left\{\lb\frac{\log n}{n}\rb^{1/3} + \Delta_{N}\right\}.\]
  \end{remark}

\subsection{Proof of Theorem \ref{thm:double_robustness_w}}
Let $\qt(\beta; F)$ denote the $\beta$-th quantile of a distribution
function $F$, i.e.
\[
\qt(\beta; F) = \inf\{z: F(z)\ge \beta\} = \sup\{z: F(z) < \beta\}.
\]
We start with two lemmas. 
\begin{lemma}[Equation (2) in Lemma 1 from \cite{barber2019conformal}]\label{lem:delta_infty}
  Let $v_{1}, \ldots, v_{n+1} \in \R$ and $(p_{1}, \ldots,
  p_{n+1}) \in \R$ be non-negative reals summing to
  $1$. Then for any $\beta \in [0, 1]$ and
  \begin{multline*}
    v_{n+1}\le \qt\lb\beta; \sum_{i=1}^{n+1}p_{i}\delta_{v_{i}}\rb
    \quad \Longleftrightarrow \quad v_{n+1}\le \qt\lb \beta; \sum_{i=1}^{n}p_{i}\delta_{v_{i}} + p_{n+1}\delta_{\infty}\rb.
  \end{multline*}
\end{lemma}

\begin{lemma}[Equation (10) from \cite{berrett2019conditional}]\label{lem:tv}
  Let $\tv(Q_{1X}, Q_{2X})$ denote the total-variation distance between $Q_{1X}$ and $Q_{2X}$. Then
  \[\tv(Q_{1X}\times P_{Y\mid X}, Q_{2X}\times P_{Y\mid X}) = \tv(Q_{1X}, Q_{2X}).\]
\end{lemma}

Returning to the proof of Theorem \ref{thm:double_robustness_w}, we first consider the case where $Q_{X}$ is absolutely continuous with respect to $P_{X}$, i.e.
  \[\P_{X \sim Q_{X}}(w(X) < \infty) = 1.\]
  In this case, for any measurable function $f$,
  \begin{equation}
    \label{eq:transformation}
    \E_{X\sim Q_{X}}[f(X)] = \E_{X\sim P_{X}}[w(X)f(X)].
  \end{equation}
  On the other hand, it always holds that $\P_{X\sim P_{X}}(w(X) < \infty) = 1$. In addition, the assumption $\E_{X\sim P_{X}}[\hat{w}(X)\mid \Z_{\tr}] < \infty$ implies that $\P_{X\sim P_{X}}(\hat{w}(X) < \infty) = 1$. By \eqref{eq:transformation},
  \[\P_{X\sim Q_{X}}(\hat{w}(X) < \infty) = 1 - \E_{X\sim P_{X}}[w(X)I(\hat{w}(X) = \infty)].\]
  Since the integrand is non-negative,
  \begin{multline*}
    \E_{X\sim P_{X}}[w(X)I(\hat{w}(X) = \infty)]\\
    = \lim_{K\rightarrow \infty}\E_{X\sim P_{X}}[w(X)I(w(X)\le K, \hat{w}(X) = \infty)]
     \le \lim_{K\rightarrow \infty} K \P_{X\sim P_{X}}(\hat{w}(X) = \infty) = 0.
  \end{multline*}
  Thus, we also have
  \[\P_{X\sim Q_{X}}(\hat{w}(X) < \infty) = 1.\]

Index the calibration fold by $\{1, \ldots, n\}$ and let
$(X_{n+1}, Y_{n+1})\sim Q_{X}\times P_{Y\mid X}$. Write $Z_{i}$ for
$(X_{i}, Y_{i})$ and $V$ for $(V_{1}, \ldots, V_{n + 1})$. For
notational convenience, we suppress the subscripts $N$ and $n$ in
$\hat{q}, \hat{w}, \hat{C}$ as well as in $\hat{p}_{i}(x)$ and $\eta(x)$.
Next, for any permutation $\pi$ on $\{1, \ldots, n + 1\}$ and
$v^{*}\in \R^{n+1}$, let
$v_{\pi}^{*} = (v_{\pi(1)}^{*}, \ldots, v_{\pi(n+1)}^{*})$. Further,
let $\ell(z)$ be the joint density of
$\Z = (Z_{1}, \ldots, Z_{n + 1})$ and $p(z)$ be the density of $Z_{1}$
(with respect to a dominating measure). Letting $\cE(v)$ denote the
unordered set of $v$, it is easy to see that
\begin{equation}
  \label{eq:weighted_conformal_cond}
  \lb V \mid \cE(V) = \cE(v^{*}), \Z_{\tr}\rb \stackrel{d}{=} v_{\Pi}^{*},
\end{equation}
where $\Pi$ is a random permutation with
\[\P\lb\Pi = \pi\mid \Z_{\tr}\rb = \frac{p(z_{\pi}^{*})}{\sum_{\pi}p(z_{\pi}^{*})} = \frac{w(X_{\pi(n + 1)})}{\sum_{\pi}w(X_{\pi(n + 1)})} = \frac{w(X_{\pi(n + 1)})}{n!\sum_{i=1}^{n+1}w(X_{i})}.\]
Note that this conditional probability is well-defined because $w(X) < \infty$ almost surely under both $P_{X}$ and $Q_{X}$. As a result, for any $j\in \{1, 2, \ldots, n + 1\}$,
\[\P\lb\Pi(n+1) = j\mid \Z_{\tr}\rb = \frac{w(X_{j})}{\sum_{i=1}^{n+1}w(X_{i})} = p_{j}(X_{n+1}),\]
where $p_{n+1}$ denotes $p_{\infty}$ for notational convenience. This
gives 
\begin{equation}
  \label{eq:Vn+1}
  \lb V_{n+1} \mid \cE(V) = \cE(v^{*}), \Z_{\tr}\rb \stackrel{d}{=} v_{\Pi(n+1)}^{*}\sim \sum_{i=1}^{n+1}p_{i}(X_{n+1})\delta_{v_{i}^{*}}.
\end{equation}
Note that $p_{i}$ involves the true likelihood ratio function $w(x)$ and thus is different from $\hat{p}_{i}$.

Let $\td{Q}_{X}$ be a measure with
\[d\td{Q}_{X}(x) = \hat{w}(x) dP_{X}(x).\]
Since $\E_{X\sim P_{X}}[\hat{w}(X)] = 1$, $\P_{X\sim P_{X}}(\hat{w}(X) < \infty) = 1$. As a result, $\td{Q}_{X}$ is a
probability measure.
Consider now a new sample
$(\td{X}_{n+1}, \td{Y}_{n+1})\sim \td{Q}_{X}\times P_{Y \mid X}$. Let
$\td{V}_{n+1}$ denote the non-conformity score of
$(\td{X}_{n+1}, \td{Y}_{n+1})$ and set
$\td{V} = (V_{1}, \ldots, V_{n}, \td{V}_{n+1})$. Using the same
argument as for \eqref{eq:Vn+1}, we have
\begin{equation}
  \label{eq:Vn+1hat}
  \lb \td{V}_{n+1} \mid \cE(\td{V})= \cE(v^{*}), \Z_{\tr}\rb \sim \sum_{i=1}^{n+1}\hat{p}_{i}(\td{X}_{n+1})\delta_{v_{i}^{*}}.
\end{equation}
Note that each $\hat{p}_{i}(\td{X}_{n+1})$,
  $i = 1,\ldots, n+1$ is well-defined since $\hat{w}(X_i)$ is almost
  surely finite under both $P_{X}$ and $Q_{X}$. As a consequence,
\begin{align}
  & \P\lb\td{Y}_{n+1}\in \hat{C}(\td{X}_{n+1})\mid \Z_{\tr}\rb\nonumber\\
  & = \P\lb \td{V}_{n + 1}\le \eta(\td{X}_{n+1})\mid \Z_{\tr}\rb\nonumber\\
  & = \P\lb \td{V}_{n + 1}\le \qt\lb 1 - \alpha; \sum_{i=1}^{n}\hat{p}_{i}(\td{X}_{n+1})\delta_{V_{i}} + \hat{p}_{\infty}(\td{X}_{n+1})\delta_{\infty}\rb\mid \Z_{\tr}\rb\nonumber\\
  & \stackrel{(i)}{=} \P\lb \td{V}_{n + 1}\le \qt\lb 1 - \alpha; \sum_{i=1}^{n}\hat{p}_{i}(\td{X}_{n+1})\delta_{V_{i}} + \hat{p}_{\infty}(\td{X}_{n+1})\delta_{\td{V}_{n+1}}\rb\mid \Z_{\tr}\rb\nonumber\\
  & = \E \P\lb \td{V}_{n + 1}\le \qt\lb 1 - \alpha; \sum_{i=1}^{n}\hat{p}_{i}(\td{X}_{n+1})\delta_{V_{i}} + \hat{p}_{\infty}(\td{X}_{n+1})\delta_{\td{V}_{n+1}}\rb \mid \cE(\td{V}), \Z_{\tr}\rb\nonumber\\
  & \stackrel{(ii)}{\ge} 1 - \alpha, \label{eq:coverage_tilde}
\end{align}
where (i) uses Lemma \ref{lem:delta_infty} and (ii) uses \eqref{eq:Vn+1hat} and the
definition of $\qt(\beta; F)$.  By Lemma \ref{lem:tv},
\[\tv\lb Q_{X}\times P_{Y\mid X}, \td{Q}_{X}\times P_{Y\mid X}\rb = \tv\lb Q_{X}, \td{Q}_{X}\rb,\]
and as a consequence,
\begin{equation}
  \label{eq:tv_Q_tdQ}
  \big|\P\lb Y_{n+1}\in \hat{C}(X_{n+1})\mid \Z_{\tr}, \Z_{\ca}\rb - \P\lb\td{Y}_{n+1}\in \hat{C}(\td{X}_{n+1})\mid \Z_{\tr}, \Z_{\ca}\rb\big|\le \tv(Q_{X}, \td{Q}_{X}),
\end{equation}
which implies that 
\[\P\lb Y_{n+1}\in \hat{C}(X_{n+1})\mid \Z_{\tr}, \Z_{\ca}\rb\ge \P\lb \td{Y}_{n+1}\in \hat{C}(\td{X}_{n+1})\mid \Z_{\tr}, \Z_{\ca}\rb - \tv(Q_{X}, \td{Q}_{X}).\]
Taking expectation over $\Z_{\ca}$, we have
\[\P\lb Y_{n+1}\in \hat{C}(X_{n+1})\mid \Z_{\tr}\rb \ge \P\lb \td{Y}_{n+1}\in \hat{C}(\td{X}_{n+1})\mid \Z_{\tr}\rb - \tv(Q_{X}, \td{Q}_{X})\ge 1 - \alpha - \tv(Q_{X}, \td{Q}_{X}).\]
Using the integral definition of total-variation distance and \eqref{eq:transformation},
\begin{align*}
  \tv(Q_{X}, \td{Q}_{X}) & = \frac{1}{2}\int |\hat{w}(x)dP_{X}(x) - dQ_{X}(x)| = \frac{1}{2}\int |\hat{w}(x)dP_{X}(x) - w(x)dP_{X}(x)|\\
  & = \frac{1}{2}\E_{X\sim P_{X}}|\hat{w}(X) - w(X)|
\end{align*}
Taking expectation over $\Z_{\tr}$, we have
\[\P\lb Y_{n+1}\in \hat{C}(X_{n+1})\rb\ge 1 - \alpha - \frac{1}{2}\E_{X\sim P_{X}}|\hat{w}(X) - w(X)|.\]
This proves \eqref{eq:unconditional_coverage_rate_w} when $\P_{X\sim Q_{X}}(w(X) < \infty) = 1$.

Next, we extend the result to the case where $\P_{X\sim Q_{X}}(w(X) < \infty) < 1$. If $\P_{X\sim P_{X}}(\hat{w}(X) < \infty) < 1$, it is clear that $\E_{X\sim P_{X}}|\hat{w}(X) - w(X)| = \infty$ and \eqref{eq:unconditional_coverage_rate_w} holds trivially. Thus, we assume $\P_{X\sim P_{X}}(\hat{w}(X) < \infty) = 1$ in the remainder.

Let $Q'_{X}$ denote the distribution $Q_{X}$ conditional on the event $E_{\infty} \triangleq \{x: w(x) < \infty\}$; that is, 
\[dQ'_{X}(x) = \frac{I(x\in
    E_{\infty})dQ_{X}(x)}{\P_{X\sim Q_{X}}(E_{\infty})}.\] Further, set
$w'(x) = dQ'_{X}(x) / dP_{X}(x)$ and
$\hat{w}'(x) = \hat{w}(x)I(x \in E_{\infty}) / \P_{X\sim Q_{X}}(E_{\infty})$. Note
that $\hat{C}(x)$ remains the same on $E_{\infty}$ when $\hat{w}$ is
replaced by $\hat{w}'$ and $Q_{X}$ is replaced by $Q'_{X}$, because
the weighted-split-CQR algorithm is invariant with respect to
rescalings of the covariate shift estimate. Since
$\P_{X\sim Q'_{X}}(w(X) < \infty) = 1$, \eqref{eq:unconditional_coverage_rate_w}
implies that
\[\P_{(X, Y)\sim Q'_{X}\times P_{Y\mid X}}\lb Y\in \hat{C}(X)\rb\ge 1 - \alpha - \frac{1}{2}\E_{X\sim P_{X}}|\hat{w}'(X) - w'(X)|.\]
It can be reformulated as
\[\P\lb Y_{n+1}\in \hat{C}(X_{n+1})\mid w(X_{n+1}) < \infty\rb\ge 1 - \alpha - \frac{1}{2\P_{X\sim Q_{X}}(E_{\infty})}\E_{X\sim P_{X}}|\hat{w}(X) - w(X)|.\]
On the other hand, when $w(X_{n+1}) = \infty$, $\eta(X_{n+1}) = \infty$, implying that $\hat{C}(X_{n+1}) = (-\infty, \infty)$. As a result,
\[\P\lb Y_{n+1}\in \hat{C}(X_{n+1})\mid w(X_{n+1}) = \infty\rb = 1.\]
Putting the two pieces together, we conclude that
\begin{align*}
  & \P\lb Y_{n+1}\in \hat{C}(X_{n+1})\rb\\
  & \begin{multlined}
    = \P\lb Y_{n+1}\in \hat{C}(X_{n+1})\mid w(X_{n+1}) < \infty\rb \P\lb w(X_{n+1}) < \infty\rb\\
    + \P\lb Y_{n+1}\in \hat{C}(X_{n+1})\mid w(X_{n+1}) = \infty\rb \P\lb w(X_{n+1}) = \infty\rb
  \end{multlined}\\
  & \ge  (1 - \alpha)\P_{X\sim Q_{X}}(E_{\infty}) + \P_{X\sim Q_{X}}(E_{\infty}^{c}) - \frac{1}{2}\E_{X\sim P_{X}}|\hat{w}(X) - w(X)|\\
  & \ge 1 - \alpha - \frac{1}{2}\E_{X\sim P_{X}}|\hat{w}(X) - w(X)|.
\end{align*}


\subsection{Proof of Theorem \ref{thm:double_robustness_q}}

We start with the following two Rosenthal-type inequalities for sums of independent random variables with finite $(1 + \delta)$-th moments.

\begin{proposition}[Theorem 3 of \cite{rosenthal1970subspaces}]\label{prop:rosenthal}
  Let $\{Z_{i}\}_{i = 1, \ldots, n}$ be independent mean-zero random
  variables. Then for any $\delta \ge 1$, there exists $L(\delta) > 0$
  that only depends on $\delta$ such that
\[\E \bigg|\sum_{i=1}^{n}Z_{i}\bigg|^{1 + \delta}\le L(\delta)\left\{\sum_{i=1}^{n}\E |Z_{i}|^{1 + \delta} + \lb\sum_{i=1}^{n}\E |Z_{i}|^{2}\rb^{(1 + \delta) / 2}\right\}.\]
\end{proposition}

\begin{proposition}[Theorem 2 of \cite{vonbahr65}]\label{prop:vonbahresseen}
 Let $\{Z_{i}\}_{i = 1, \ldots, n}$ be independent mean-zero random variables. Then for any $\delta \in [0, 1)$,
\[\E \bigg|\sum_{i=1}^{n}Z_{i}\bigg|^{1 + \delta}\le 2\sum_{i=1}^{n}\E |Z_{i}|^{1 + \delta}.\]
\end{proposition}


For notational convenience, we suppress the subcripts $N$ and $n$ in $\hat{q}, \hat{w}, \hat{C}$ as well as in $\hat{p}_{i}(x)$ and $\eta(x)$. Note that Assumption (3) implies that $w(X)$ is almost surely finite under $Q_{X}$ and $\hat{w}(X)$ is almost surely finite under $P_{X}$. By the same reasoning as in the proof of \eqref{eq:unconditional_coverage_rate_w}, $w(X)$ is almost surely finite under $P_{X}$ and $\hat{w}(X)$ is almost surely finite under $Q_{X}$. 

Let $\eps < r / 2$ and $(\td{X}, \td{Y})$ denote a generic random vector drawn from
$Q_{X}\times P_{Y\mid X}$, which is independent of the data. Then
\begin{align}
  &\P(\td{Y} \in \hat{C}(\td{X}) \mid \td{X})\nonumber\\
  & = \P\lb\max\{\hat{q}_{\alpha_{\lo}}(\td{X}) - \td{Y}, \td{Y} - \hat{q}_{\alpha_{\hi}}(\td{X})\}\le \eta(\td{X})\mid \td{X}\rb\nonumber\\
  &\ge \P\lb\max\{q_{\alpha_{\lo}}(\td{X}) - \td{Y}, \td{Y} - q_{\alpha_{\hi}}(\td{X})\}\le \eta(\td{X}) - H(\td{X})\mid \td{X}\rb\nonumber\\
  & \ge \P\lb\max\{q_{\alpha_{\lo}}(\td{X}) - \td{Y}, \td{Y} - q_{\alpha_{\hi}}(\td{X})\}\le -\eps - H(\td{X})\mid \td{X}\rb - \P(\eta(\td{X}) < -\eps\mid \td{X})\nonumber\\
  & \ge \P\lb\max\{q_{\alpha_{\lo}}(\td{X}) - \td{Y}, \td{Y} - q_{\alpha_{\hi}}(\td{X})\}\le -\eps - H(\td{X})I(H(\td{X})\le \eps)\mid \td{X}\rb\nonumber\\
  & \quad - I(H(\td{X}) > \eps) - \P(\eta(\td{X}) < -\eps\mid \td{X})\nonumber\\
  & \stackrel{(i)}{\ge}\P\lb\max\{q_{\alpha_{\lo}}(\td{X}) - \td{Y}, \td{Y} - q_{\alpha_{\hi}}(\td{X})\}\le 0 \mid \td{X}\rb - b_{2}\left\{\eps + H(\td{X})I(H(\td{X})\le \eps)\right\}\nonumber\\
  & \quad - I(H(\td{X}) > \eps) - \P(\eta(\td{X}) < -\eps\mid \td{X})\nonumber\\
  & \ge \P\lb\max\{q_{\alpha_{\lo}}(\td{X}) - \td{Y}, \td{Y} - q_{\alpha_{\hi}}(\td{X})\}\le 0 \mid \td{X}\rb - b_{2}\left\{\eps + H(\td{X})\right\}\nonumber\\
  & \quad - I(H(\td{X}) > \eps) - \P(\eta(\td{X}) < -\eps \mid \td{X})\nonumber\\
  & \stackrel{(ii)}{=} 1 - \alpha - b_{2}\left\{\eps + H(\td{X})\right\} - I(H(\td{X}) > \eps) - \P(\eta(\td{X}) < -\eps \mid \td{X}); \label{eq:conditional_coverage}
\end{align}
above, (i) uses the condition that $2\eps < r$ and the assumption
(2), and (ii) follows from the assumption (1) and the definitions of $q_{\alpha_{\lo}}$ and $q_{\alpha_{\hi}}$ that $\P(\td{Y}\in [q_{\alpha_{\lo}}(\td{X}), q_{\alpha_{\hi}}(\td{X})]) = \alpha_{\hi} - \alpha_{\lo} = 1 - \alpha$. 

Next, we derive an upper bound on
  $\P(\eta(\td{X}) < -\eps \mid \td{X})$. Let $G$ denote the
  cumulative distribution function of the random distribution
  $\sum_{i=1}^{n}\hat{p}_{i}(\td{X})\delta_{V_{i}} +
  \hat{p}_{\infty}(\td{X})\delta_{\infty}$. Again, $G$ implicitly
  depends on $N$, $n$ and $\td{X}$. Then $\eta(\td{X}) < -\eps$
  implies $G(-\eps)\ge 1 - \alpha$, and thus,
\[\P\lb\eta(\td{X}) < -\eps\mid \td{X}\rb\le \P\lb G(-\eps)\ge 1 - \alpha \mid \td{X}\rb, \,\, \text{a.s.}.\]
Let $G^{*}(-\eps)$ denote the expectation of $G(-\eps)$ conditional on
$\D = \{\Z_{\tr}, (X_{i})_{i=1}^{n}, \td{X}\}$, namely,
\[G^{*}(-\eps) = \E[G(-\eps)\mid \D] = \sum_{i=1}^{n}\hat{p}_{i}(\td{X})\P(V_{i}\le -\eps\mid \D).\]
For any $t > 0$, the triangle inequality implies that
\begin{equation}
  \label{eq:etaX1}
  \P\lb\eta(\td{X}) < -\eps\mid \td{X}\rb\le \P\lb G(-\eps) - G^{*}(-\eps)\ge t \mid \td{X}\rb + \P\lb G^{*}(-\eps)\ge 1 - \alpha - t \mid \td{X}\rb, \,\, \text{a.s.}.
\end{equation}
To bound the first term, we note that
\[G(-\eps) - G^{*}(-\eps) = \sum_{i=1}^{n}\hat{p}_{i}(\td{X})\lb
  I(V_{i}\le - \eps) - \P(V_{i}\le -\eps\mid \D)\rb.\] Conditional on
$\D$, $G(-\eps) - G^{*}(-\eps)$ is sub-Gaussian with
parameter
\[\hat{\sigma}^{2} = \sum_{i=1}^{n}\hat{p}_{i}(\td{X})^{2}.\]
For any $t > 0$,
\[\P\lb G(-\eps) - G^{*}(-\eps) \ge t\mid \D\rb\le \exp\lb
  -\frac{t^{2}}{2\hat{\sigma}^{2}}\rb.\]
Let $\gamma_{n}$ be any fixed sequence with $\gamma_{n} = O(1)$. Taking expectation over $\D\setminus \{\td{X}\}$, we obtain that
\begin{align*}
  &\P\lb G(-\eps) - G^{*}(-\eps) \ge t \mid \td{X}\rb\\
  & \le \E\left[\exp\lb -\frac{t^{2}}{2\hat{\sigma}^{2}}\rb\mid \td{X}\right] \\
  & \le \exp\lb - \frac{t^2}{2\gamma_{n}}\rb + \P\lb\hat{\sigma}^2 \ge \gamma_{n}\mid \td{X}\rb\\
  & = \exp\lb - \frac{t^2}{2\gamma_{n}}\rb + \P\lb \frac{\sum_{i=1}^{n}\hat{w}(X_i)^2}{\lb\sum_{i=1}^{n}\hat{w}(X_i) + \hat{w}(\td{X})\rb^2}\ge \gamma_{n}\mid \td{X}\rb\\
  & \le \exp\lb - \frac{t^2}{2\gamma_{n}}\rb + \P\lb \frac{\sum_{i=1}^{n}\hat{w}(X_i)^2}{\lb\sum_{i=1}^{n}\hat{w}(X_i)\rb^2}\ge \gamma_{n}\mid \td{X}\rb\\
  & \stackrel{(i)}{=} \exp\lb - \frac{t^2}{2\gamma_{n}}\rb + \P\lb \frac{\sum_{i=1}^{n}\hat{w}(X_i)^2}{\lb\sum_{i=1}^{n}\hat{w}(X_i)\rb^2}\ge \gamma_{n}\rb\\
  & \le \exp\lb - \frac{t^2}{2\gamma_{n}}\rb + \P\lb\sum_{i=1}^{n}\hat{w}(X_i)\le \frac{n}{2}\rb + \P\lb\sum_{i=1}^{n}\hat{w}(X_i)^2\ge \frac{n^2\gamma_{n}}{4}\rb\\
  & \le \exp\lb - \frac{t^2}{2\gamma_{n}}\rb + \P\lb\sum_{i=1}^{n}|\hat{w}(X_i) - 1|\ge \frac{n}{2}\rb + \P\lb\sum_{i=1}^{n}\hat{w}(X_i)^2\ge \frac{n^2\gamma_{n}}{4}\rb,
\end{align*}
where (i) uses the fact that $\td{X}$ is independent of
$(\hat{w}(X_i))_{i=1}^{n}$. Note that this bound holds uniformly with $\td{X}$. Throughout the rest of the proof, we write
$a_{1n}\preceq a_{2n}$ if there exists a constant $\const$ that only
depends on $r, b_1, b_2, \delta, M, k, \ell$ such that
$a_{1n}\le \const a_{2n}$ for all $n$. We consider two cases:
\begin{enumerate}[(1)]
\item If $\delta \ge 1$, then by Markov's inequality,
  \[\P\lb\sum_{i=1}^{n}\hat{w}(X_i)^2\ge \frac{n^2\gamma_{n}}{4}\rb\le \frac{4\E[\sum_{i=1}^{n}\hat{w}(X_i)^2]}{n^2\gamma_{n}} = \frac{4\E[\hat{w}(X_1)^2]}{n\gamma_{n}}\preceq \frac{1}{n\gamma_{n}}.\]
  Since $\E[\hat{w}(X_i)\mid \Z_\tr] = 1$, we have $\E[\hat{w}(X_i)] = 1$. By Markov's inequality and Proposition \ref{prop:rosenthal},
  \begin{align}
    &\P\lb \sum_{i=1}^{n}|\hat{w}(X_i) - 1|\ge \frac{n}{2}\rb \nonumber\\
    & \le \frac{2^{1 + \delta}}{n^{1 + \delta}}\E\lb\sum_{i=1}^{n}|\hat{w}(X_i) - \E[\hat{w}(X_i)]|\rb^{1 + \delta}\nonumber\\
    & \preceq \frac{1}{n^{1 + \delta}}\left\{ n\E|\hat{w}(X_i) - \E[\hat{w}(X_i)]|^{1 + \delta} + n^{(1 + \delta) / 2}\lb\E|\hat{w}(X_i) - \E[\hat{w}(X_i)]|^{2}\rb^{(1 + \delta) / 2}\right\}\nonumber\\
    & \stackrel{(i)}{\preceq} \frac{1}{n^{1 + \delta}}\left\{ n\E|\hat{w}(X_i)|^{1 + \delta} + n^{(1 + \delta) / 2}(\E|\hat{w}(X_i)|^2)^{(1 + \delta) / 2}\right\}\nonumber\\
    & \preceq \frac{1}{n^{(1 + \delta) / 2}},\label{eq:denom_case1}
  \end{align}
  where (i) follows from the H\"{o}lder's
  inequality which gives 
  \[\E|\hat{w}(X_i) - \E[\hat{w}(X_i)]|^{1 + \delta}\le 2^{\delta}\lb \E|\hat{w}(X_i)|^{1+\delta} + |\E[\hat{w}(X_i)]|^{1+\delta}\rb\le 2^{1 + \delta}\E|\hat{w}(X_i)|^{1+\delta}.\]
  Piecing things together yields
  \[\P\lb G(-\eps) - G^{*}(-\eps) \ge t \mid \td{X}\rb\preceq \exp\lb - \frac{t^2}{2\gamma_{n}}\rb + \frac{1}{n^{(1 + \delta) / 2}} + \frac{1}{n\gamma_{n}} \preceq \exp\lb - \frac{t^2}{2\gamma_{n}}\rb + \frac{1}{n\gamma_{n}},\]
  where the last step follows from the fact that $\delta\ge 1$ and $\gamma_{n} = O(1)$.
\item If $\delta < 1$, then by Markov's inequality,
  \begin{align*}
    &\P\lb\sum_{i=1}^{n}\hat{w}(X_i)^2\ge \frac{n^2\gamma_{n}}{4}\rb\le \frac{\E\left[\lb\sum_{i=1}^{n}\hat{w}(X_i)^2\rb^{(1 + \delta) / 2}\right]}{(n^2\gamma_{n})^{(1 + \delta) / 2}}
  \le \frac{\E\left[\sum_{i=1}^{n}\hat{w}(X_i)^{1 + \delta}\right]}{(n^2\gamma_{n})^{(1 + \delta) / 2}}\preceq \frac{1}{n^{\delta}\gamma_{n}^{(1 + \delta)/2}},
  \end{align*}
  where the last step follows from the simple fact that $\|x\|_{p}\le \|x\|_{1}$ for $p\ge 1$, with $p = 2 / (1 + \delta)$ and $x_i = \hat{w}(X_i)^{1 + \delta}$. By Markov's inequality and Proposition \ref{prop:vonbahresseen},
  \begin{multline}
    \P\lb \sum_{i=1}^{n}|\hat{w}(X_i) - 1|\ge \frac{n}{2}\rb \\
    \le \frac{2^{1 + \delta}}{n^{1 + \delta}}\E\lb\sum_{i=1}^{n}|\hat{w}(X_i) - \E[\hat{w}(X_i)]|\rb^{1 + \delta} \preceq \frac{2n\E|\hat{w}(X_i) - \E[\hat{w}(X_i)]|^{1 + \delta}}{n^{1 + \delta}} \preceq \frac{1}{n^{\delta}}.\label{eq:denom_case2}
  \end{multline}
  Piecing things together yields 
  \[\P\lb G(-\eps) - G^{*}(-\eps) \ge t \mid \td{X}\rb\preceq \exp\lb - \frac{t^2}{2\gamma_{n}}\rb + \frac{1}{n^{\delta}} + \frac{1}{n^{\delta}\gamma_{n}^{(1 + \delta) / 2}} \preceq \exp\lb - \frac{t^2}{2\gamma_{n}}\rb + \frac{1}{n^{\delta}\gamma_{n}^{(1 + \delta) / 2}},\]
  where the last step follows from $\gamma_{n} = O(1)$.
\end{enumerate}
In all cases, 
\begin{equation}
  \label{eq:etaX2}
  \P\lb G(-\eps) - G^{*}(-\eps) \ge t \mid \td{X}\rb\preceq 
\exp\lb - \frac{t^2}{2\gamma_{n}}\rb + \frac{1}{n^{\delta'}\gamma_{n}^{(1 + \delta') / 2}}, \quad \delta' = \min\{\delta, 1\}.
\end{equation}

~\\
\noindent Next, we almost surely bound the term
$\P\lb G^{*}(-\eps)\ge 1 - \alpha - t \mid \td{X}\rb$. By the triangle
inequality and definition of $H(\cdot)$,
\[V_{i}\ge \max\{q_{\alpha_{\lo}}(X_{i}) - Y_{i}, Y_{i} - q_{\alpha_{\hi}}(X_{i})\} - H(X_{i})\triangleq V_{i}^{*} - H(X_{i}).\]
By Assumptions (1) and (2), $\P(V_{i}^{*}\le 0\mid \D) = \alpha_{\hi} - \alpha_{\lo} = 1 - \alpha$. Conditioning on $\D$, $H(X_{i})$ is deterministic. Since $\eps < r / 2 < 2r$,
\begin{align}
  G^{*}(-\eps)
  &\le \sum_{i=1}^{n}\hat{p}_{i}(\td{X})\left\{ I\lb H(X_{i})\ge \frac{\eps}{2}\rb + \P\lb V_{i}^{*}\le -\frac{\eps}{2}\mid \D\rb\right\}\nonumber\\
  & \le \sum_{i=1}^{n}\hat{p}_{i}(\td{X})\left\{ I\lb H(X_{i})\ge \frac{\eps}{2}\rb + \P\lb V_{i}^{*}\le -\frac{\eps}{2}\mid \D\rb\right\}\nonumber\\
  & \le \sum_{i=1}^{n}\hat{p}_{i}(\td{X})\left\{ I\lb H(X_{i})\ge \frac{\eps}{2}\rb + \P\lb V_{i}^{*}\le 0\mid \D\rb - \frac{\eps b_{1}}{2}\right\}\nonumber\\
  & = \sum_{i=1}^{n}\hat{p}_{i}(\td{X})\left\{ I\lb H(X_{i})\ge \frac{\eps}{2}\rb + 1 - \alpha - \frac{\eps b_{1}}{2}\right\}\nonumber\\
  & \le 1 - \alpha - \frac{\eps b_{1}}{2} + \sum_{i=1}^{n}\hat{p}_{i}(\td{X})I\lb H(X_{i})\ge \frac{\eps}{2}\rb,\label{eq:Gstar1}
\end{align}
where the last step follows from the fact that $\sum_{i=1}^{n}\hat{p}_{i}(\td{X})\le 1$. If $t \le \eps b_1 / 4$, \eqref{eq:Gstar1} implies that 
\begin{align*}
  &\P\lb G^{*}(-\eps)\ge 1 - \alpha - t \mid \td{X}\rb\\
  & \le \P\lb\sum_{i=1}^{n}\hat{p}_{i}(\td{X})I\lb H(X_i)\ge \frac{\eps}{2}\rb\ge \frac{\eps b_1}{2} - t\mid \td{X}\rb\\
  & = \P\lb\frac{\sum_{i=1}^{n}\hat{w}(X_i)I(H(X_i)\ge \eps / 2)}{\sum_{i=1}^{n}\hat{w}(X_i) + \hat{w}(\td{X})}\ge \frac{\eps b_1}{2} - t\mid \td{X}\rb\\
  & \le \P\lb\frac{\sum_{i=1}^{n}\hat{w}(X_i)I(H(X_i)\ge \eps / 2)}{\sum_{i=1}^{n}\hat{w}(X_i)}\ge \frac{\eps b_1}{2} - t\mid \td{X}\rb\\
  & \stackrel{(i)}{=} \P\lb\frac{\sum_{i=1}^{n}\hat{w}(X_i)I(H(X_i)\ge \eps / 2)}{\sum_{i=1}^{n}\hat{w}(X_i)}\ge \frac{\eps b_1}{2} - t\rb\\
  & \le \P\lb\sum_{i=1}^{n}\hat{w}(X_i)\le \frac{n}{2}\rb + \P\lb \sum_{i=1}^{n}\hat{w}(X_i)I\lb H(X_i)\ge \frac{\eps}{2}\rb\ge \frac{n(\eps b_1 - 2t)}{4}\rb\\
  & \stackrel{(ii)}{\le} \P\lb\sum_{i=1}^{n}|\hat{w}(X_i) - 1|\ge \frac{n}{2}\rb + \P\lb \sum_{i=1}^{n}\hat{w}(X_i)I\lb H(X_i)\ge \frac{\eps}{2}\rb\ge \frac{n\eps b_1}{8}\rb,
\end{align*}
where (i) follows from the independence between $\td{X}$ and $\D\setminus\{\td{X}\}$ and (ii) follows from the fact that $t \le \eps b_1 / 4$. By \eqref{eq:denom_case1} and \eqref{eq:denom_case2}, we have that
\[\P\lb\sum_{i=1}^{n}|\hat{w}(X_i) - 1|\ge \frac{n}{2}\rb\preceq \frac{1}{n^{(\delta + \delta') / 2}},\]
where $\delta' = \min\{\delta, 1\}$. By Markov's inequality,
\begin{align*}
  \P\lb \sum_{i=1}^{n}\hat{w}(X_i)I\lb H(X_i)\ge \frac{\eps}{2}\rb\ge \frac{n\eps b_1}{8}\rb &\preceq \frac{1}{\eps}\E\left[\hat{w}(X)I\lb H(X)\ge \frac{\eps}{2}\rb\right]\preceq \frac{\E[\hat{w}(X)H^{k}(X)]}{\eps^{1+k}},
\end{align*}
where the last step uses the simple fact that $I(H(X_i)\ge \eps / 2)\le (2/\eps)^{k}H^{k}(X_i)$. Therefore, for any $t\le \eps b_1 / 4$, we obtain an almost sure bound of the form 
\begin{equation}
  \label{eq:etaX3}
  \P\lb G^{*}(-\eps)\ge 1 - \alpha - t \mid \td{X}\rb \preceq \frac{1}{n^{(\delta + \delta') / 2}} + \frac{\E[\hat{w}(X)H^{k}(X)]}{\eps^{1+k}}.
\end{equation}

~\\
Combining \eqref{eq:etaX1}, \eqref{eq:etaX2} and \eqref{eq:etaX3} together and setting $t = \eps b_1 / 4$, we obtain that for any sequence $\gamma_{n} = O(1)$, 
\begin{align}
  \P\lb\eta(\td{X}) < -\eps\mid \td{X}\rb &\preceq \exp\lb - \frac{b_1^2}{32}\frac{\eps^2}{\gamma_{n}}\rb + \frac{1}{n^{\delta'}\gamma_{n}^{(1 + \delta') / 2}} + \frac{1}{n^{(\delta + \delta') / 2}} + \frac{\E[\hat{w}(X)H^{k}(X)]}{\eps^{1+k}}\nonumber\\
  & \preceq \exp\lb - \frac{b_1^2}{32}\frac{\eps^2}{\gamma_{n}}\rb + \frac{1}{n^{\delta'}\gamma_{n}^{(1 + \delta') / 2}} + \frac{\E[\hat{w}(X)H^{k}(X)]}{\eps^{1+k}}.  \label{eq:etaX4}
\end{align}
Substitute $\eps$ with $\eps_{n}$ and assume $\eps_{n} \le r / 2$
(recall the beginning of the proof). Set
\begin{equation}
  \label{eq:gamman}
  \gamma_{n} = \frac{32}{b_1^2}\frac{\eps_{n}^2}{\log n}.
\end{equation}
Clearly, $\gamma_{n} = o(1)$. Then the first term of \eqref{eq:etaX4} is $1 / n$, and thus, 
\[\P\lb\eta(\td{X}) < -\eps_{n}\mid \td{X}\rb \preceq \frac{(\log n)^{(1 + \delta') / 2}}{n^{\delta'}\eps_{n}^{1 + \delta'}} + \frac{\E[\hat{w}(X)H^{k}(X)]}{\eps_{n}^{1+k}}.\]
Equivalently, there exists a constant $\const$ that only depends on $r, b_1, b_2, \delta, M, k, \ell$, such that
\[\P\lb\eta(\td{X}) < -\eps_{n}\mid \td{X}\rb \le \const\left\{\frac{(\log n)^{(1 + \delta') / 2}}{n^{\delta'}\eps_{n}^{1 + \delta'}} + \frac{\E[\hat{w}(X)H^{k}(X)]}{\eps_{n}^{1+k}}\right\}, \,\, \text{a.s.}.\]
Together with \eqref{eq:conditional_coverage}, it implies that
\begin{multline*}
  \P(\td{Y} \in \hat{C}(\td{X}) \mid \td{X})\\
  \ge  1 - \alpha - b_{2}(\eps_{n} + H(\td{X})) - I(H(\td{X}) > \eps_{n}) - \const\left\{\frac{(\log n)^{(1 + \delta') / 2}}{n^{\delta'}\eps_{n}^{1 + \delta'}} + \frac{\E[\hat{w}(X)H^{k}(X)]}{\eps_{n}^{1+k}}\right\},
\end{multline*}
almost surely. Assume $\const\ge 2b_2$ without loss of generality. Then
\begin{multline}
  \P\lb\P(\td{Y} \in \hat{C}(\td{X}) \mid \td{X}) \le 1 - \alpha - \const\left\{\eps_{n} + \frac{(\log n)^{(1 + \delta') / 2}}{n^{\delta'}\eps_{n}^{1 + \delta'}} + \frac{\E[\hat{w}(X)H^{k}(X)]}{\eps_{n}^{1+k}}\right\}\rb\\
  \quad \le \P(H(\td{X}) > \eps_{n}).\label{eq:conditional_coverage2}
\end{multline}
For any $\beta\in (0, 1)$, let
\[\eps_{n} = \frac{(\log n)^{(1 + \delta') / 2(2 + \delta')}}{n^{\delta' / (2 + \delta')}} + \lb\E[\hat{w}(X)H^{k}(X)]\rb^{1 / (2+k)} + \frac{\lb\E[w(X)H^{\ell}(X)]\rb^{1/\ell}}{\beta^{1/\ell}}.\]
Then
\begin{equation}
  \label{eq:eps_cond}
  \frac{(\log n)^{(1 + \delta') / 2}}{n^{\delta'}\eps_{n}^{1 + \delta'}}, \frac{\E[\hat{w}(X)H^{k}(X)]}{\eps_{n}^{1+k}}\le \eps_{n},
\end{equation}
and by Markov's inequality and \eqref{eq:transformation},
\[\P(H(\td{X}) > \eps_{n})\le \frac{\E[H^{\ell}(\td{X})]}{\eps_{n}^{\ell}} = \frac{\E[w(X)H^{\ell}(X)]}{\eps_{n}^{\ell}}\le \beta.\]
Furthermore, the Assumption (4) implies that $\eps_{n}\le r / 2$ when $N\ge N(r)$ and $n\ge n(r)$ for some constants $N(r), n(r)$ that only depend on $r$. Replacing $\const$ by $3\const$, we obtain that, for $N\ge N(r), n\ge n(r)$, 
\[\P\lb\P(\td{Y} \in \hat{C}(\td{X}) \mid \td{X}) \le 1 - \alpha - \const\eps_{n}\rb\le \beta.\]
We can further enlarge $\const$ so that $B\eps_{n} \ge 1 - \alpha$ when $N < N(r)$ or $n < n(r)$, in which case \eqref{eq:conditional_coverage_rate_q} trivially holds.

~\\
\noindent To prove the unconditional result, we note that \eqref{eq:conditional_coverage2} implies
\[\P(\td{Y} \in \hat{C}(\td{X}))\ge \lb 1 - \alpha - \const\left\{\eps_{n} + \frac{(\log n)^{(1 + \delta') / 2}}{n^{\delta'}\eps_{n}^{1 + \delta'}} + \frac{\E[\hat{w}(X)H^{k}(X)]}{\eps_{n}^{1+k}}\right\}\rb(1 - \P(H(\td{X}) > \eps_{n})).\]
Let
\[\eps_{n} = \frac{(\log n)^{(1 + \delta') / 2(2 + \delta')}}{n^{\delta' / (2 + \delta')}} + \lb\E[\hat{w}(X)H^{k}(X)]\rb^{1 / (2+k)} + \lb\E[w(X)H^{\ell}(X)]\rb^{1/ (1+\ell)}.\]
Then \eqref{eq:eps_cond} remains to hold. 
By Markov's inequality and \eqref{eq:transformation},
\[\P(H(\td{X}) > \eps_{n})\le \frac{\E[H^{\ell}(\td{X})]}{\eps_{n}^{\ell}} = \frac{\E[w(X)H^{\ell}(X)]}{\eps_{n}^{\ell}}\le \eps_{n}.\]
Furthermore, Assumption (4) implies that $\eps_{n}\le r / 2$ when $N$ and $n$ are sufficiently large, in which case,
\[\P(\td{Y} \in \hat{C}(\td{X}))\ge \lb 1 - \alpha - 3\const\eps_{n}\rb(1 - \eps_{n})\ge 1 - \alpha - (3\const+1)\eps_{n}.\]
Similar to \eqref{eq:conditional_coverage_rate_q}, we can enlarge the constant to make \eqref{eq:unconditional_coverage_rate_q} hold 
when $N$ or $n$ is not sufficiently large.

  \subsection{An asymptotic result}\label{subsec:asymptotic}

  Theorem \ref{thm:double_robustness_w} and Theorem \ref{thm:double_robustness_q} together imply the following asymptotic
  result, which is a generalization of Theorem
  \ref{thm:double_robustness_ATE} from Section
  \ref{subsec:double_robustness}.
\begin{corollary}\label{cor:double_robustness}
With the same notation as in Theorem \ref{thm:double_robustness_w}, assume
  that either B1 or B2 (or both) is satisfied:
  \begin{enumerate}[\textbf{B}1]
  \item $\displaystyle \lim_{N\rightarrow \infty}\E |\hat{w}_{N}(X) - w(X)| = 0$;
  \item the conditions (1)-(4) in Theorem \ref{thm:double_robustness_q} hold.
  \end{enumerate}
  Then
  \begin{equation}
    \label{eq:unconditional_coverage_general}
    \lim_{N, n\rightarrow\infty}\P_{(X, Y)\sim Q_{X}\times P_{Y\mid X}}(Y \in \hat{C}_{N, n}(X))\ge 1 - \alpha.
  \end{equation}
  Furthermore, under \textbf{B}2, for any $\eps > 0$,
    \begin{equation}
    \label{eq:conditional_coverage_general}
    \lim_{N, n\rightarrow\infty}\P_{X\sim Q_{X}}\lb \P(Y\in \hat{C}_{N, n}(X)\mid X)\le 1 - \alpha - \eps\rb = 0.
  \end{equation}
\end{corollary}

\section{Proofs of Other Results}\label{app:miscellaneous}
\subsection{Proof of Proposition \ref{prop:equal}}

Note that it remains to prove the general result since (1) and (2) are special cases. The lower bound is proved by \eqref{eq:unconditional_coverage_rate_w}. For the upper bound, we first note that $\E[\hat{w}(X)^{\r}] < \infty$ implies that $\P_{X\sim P_{X}}(\hat{w}(X) < \infty) = 1$ and $\E[\hat{w}(X)] < \infty$. Thus, we can assume $\E[\hat{w}(X)] = 1$ without loss of generality due to the invariance to rescalings of weighted split-CQR. Let $\td{Q}_{X}$ be a probability measure with $d\td{Q}_{X} = \hat{w}(X)dP_{X}$ and $(\td{X}, \td{Y})$ be a sample from $\td{Q}_{X}\times P_{Y\mid X}$ that is independent of the data. By H\"{o}lder's inequality, 
\[\E [\hat{w}(\td{X})] = \int \frac{d\td{Q}_{X}}{dP_{X}}d\td{Q}_{X} = \E_{X\sim P_{X}}[\hat{w}(X)^{2}] \le M_{\r}^2 < \infty,\]
By \eqref{eq:coverage_tilde} again with $(\td{X}, \td{Y})$ denoting $(\td{X}_{n+1}, \td{Y}_{n+1})$ for simplicity,
\begin{align*}
  &\P\lb\td{Y}\in \hat{C}(\td{X})\mid \Z_{\tr}\rb\\
  & = \E \P\lb \td{V}_{n + 1}\le \qt\lb 1 - \alpha; \sum_{i=1}^{n}\hat{p}_{i}(\td{X})\delta_{V_{i}^{*}} + \hat{p}_{\infty}(\td{X})\delta_{\td{V}_{n+1}}\rb \mid \cE(\td{V}), \Z_{\tr}\rb\nonumber\\
 & \le \E \lb 1 - \alpha + \max_{i\in [n]\cup \{\infty\}}\hat{p}_{i}(\td{X})\rb.
\end{align*}
 Let $\A$ denote the event that
 \[\sum_{i=1}^{n}\hat{w}(X_{i}) \le \frac{n}{2}.\]
 By \eqref{eq:denom_case1} with $\delta = 1$, we have that
 \[\P\lb\sum_{i=1}^{n}\hat{w}(X_i)\le \frac{n}{2}\rb \le \frac{c_{1}M_{\r}^{2}}{n},\]
 where $c_1$ is an absolute constant. Note that
 \[\max_{i\in [n]\cup \{\infty\}}\hat{p}_{i}(\td{X}) = \frac{\max\{\hat{w}(\td{X}), \max_{i}\hat{w}(X_{i})\}}{\hat{w}(\td{X}) + \sum_{i=1}^{n}\hat{w}(X_{i})}\le 1.\]
 Then
 \begin{align*}
   &\E \left[\frac{\max\{\hat{w}(\td{X}), \max_{i}\hat{w}(X_{i})\}}{\hat{w}(\td{X}) + \sum_{i=1}^{n}\hat{w}(X_{i})}\right]\\
   & \le \E\left[\frac{\max\{\hat{w}(\td{X}), \max_{i}\hat{w}(X_{i})\}}{\hat{w}(\td{X}) + \sum_{i=1}^{n}\hat{w}(X_{i})}I_{\A^{c}}\right] + \P(\A)\\
   & \le \E\left[\frac{2\max\{\hat{w}(\td{X}), \max_{i}\hat{w}(X_{i})\}}{n}I_{\A^{c}}\right] + \frac{c_{1}M_{\r}^{2}}{n}\\
   & \le \frac{2}{n}\lb \E \hat{w}(\td{X}) + \E \max_{i}\hat{w}(X_{i})\rb + \frac{c_{1}M_{\r}^{2}}{n}\\
   & \le \frac{2}{n}\lb \E \hat{w}(\td{X}) + \lb\E\sum_{i=1}^{n}\hat{w}(X_{i})^{\r}\rb^{1/\r}\rb + \frac{c_{1}M_{\r}^{2}}{n}\\
   & \le \frac{2}{n}\lb M_{\r}^{2} + n^{1 / \r}M_{\r}\rb + \frac{c_{1}M_{\r}^{2}}{n}.
 \end{align*}
This implies that
 \[\P_{(X, Y)\sim \td{Q}_{X}\times P_{Y\mid X}}(Y\in \hat{C}(X))\le 1 - \alpha + cn^{1/\r - 1}\]
 for some constant $c$ that only depends on $M_{\r}$ and $\r$. The upper bound is then proved by \eqref{eq:tv_Q_tdQ} and the same steps following \eqref{eq:tv_Q_tdQ}.

\subsection{Proof of Theorem \ref{thm:double_robustness_ATE}}
Since $\E[1 / \hat{e}_{N}(X)\mid \Z_{\tr}] < \infty$ and
$\E[1 / e(X)] < \infty$, we can here set
$$\hat{w}_{N}(x) = \frac{1 / \hat{e}_{N}(x)}{\E[1 / \hat{e}_{N}(X)\mid
  \Z_{\tr}]}, \,\, w(x) = \frac{dP_{X}(x)}{dP_{X\mid T=1}(x)} = \frac{1 / e(x)}{\E[1 / e(X)]}$$
$$\Longrightarrow \E[\hat{w}_{N}(X)\mid \Z_{\tr}] = 1 = \E[w(X)].$$ 
Thus, the assumption
\textbf{B}1 of Corollary \ref{cor:double_robustness} reduces to
\[\lim_{N\rightarrow \infty}\E\bigg|\frac{1 / \hat{e}_{N}(X)}{\E[1 / \hat{e}_{N}(X)\mid \Z_{\tr}]} - \frac{1 / e(X)}{\E[1 / e(X)]}\bigg| = 0,\]
and the assumptions \textbf{B}2 (3)-(4) of Corollary \ref{cor:double_robustness} reduce to
\[\limsup_{N\rightarrow\infty}\frac{\E [1 / \hat{e}(X)^{1 + \delta}]}{\lb\E [1 / \hat{e}(X)]\rb^{1 + \delta}} < \infty, \quad \lim_{N\rightarrow \infty}\frac{\E[H_{N}(X) / \hat{e}_{N}(X)]}{\E[1 / \hat{e}_{N}(X)]} = \lim_{N\rightarrow \infty}\frac{\E[H_{N}(X) / e(X)]}{\E[1 / e(X)]} = 0.\]
Clearly, \textbf{A}2 implies \textbf{B}2 since
$e(x), \hat{e}_{N}(x)\in [0, 1]$. Now we prove that \textbf{A}1
implies \textbf{B}1. In fact,
\begin{align*}
  &\lim_{N\rightarrow \infty}\E\bigg|\frac{1 / \hat{e}_{N}(X)}{\E[1 / \hat{e}_{N}(X)\mid \Z_{\tr}]} - \frac{1 / e(X)}{\E[1 / e(X)]}\bigg| \\
  \le & \limsup_{N\rightarrow \infty}\frac{1}{\E[1 / \hat{e}_{N}(X)\mid \Z_{\tr}]}\E\bigg|\frac{1}{\hat{e}_{N}(X)} - \frac{1}{e(X)}\bigg| + \limsup_{N\rightarrow \infty}\E \left[\frac{1}{e(X)}\right]\E\bigg|\frac{1}{\E[1 / \hat{e}_{N}(X)\mid \Z_{\tr}]} - \frac{1}{\E[1 / e(X)]}\bigg|\\
  \stackrel{(i)}{\le} & \limsup_{N\rightarrow \infty}\E\bigg|\frac{1}{\hat{e}_{N}(X)} - \frac{1}{e(X)}\bigg| + \limsup_{N\rightarrow \infty}\E \left[\frac{1}{e(X)}\right]\E \bigg|\E\left[\frac{1}{\hat{e}_{N}(X)}\mid \Z_{\tr}\right] - \E\left[\frac{1}{e(X)}\right]\bigg|\\
  \le & \limsup_{N\rightarrow \infty}\E\bigg|\frac{1}{\hat{e}_{N}(X)} - \frac{1}{e(X)}\bigg| + \limsup_{N\rightarrow \infty}\E \left[\frac{1}{e(X)}\right]\E\lb \E\left[\bigg|\frac{1}{\hat{e}_{N}(X)} - \frac{1}{e(X)}\bigg|\mid \Z_{\tr}\right]\rb\\
  \stackrel{(ii)}{=} & \lb 1 + \E \left[\frac{1}{e(X)}\right]\rb\limsup_{N\rightarrow \infty}\E\bigg|\frac{1}{\hat{e}_{N}(X)} - \frac{1}{e(X)}\bigg|;
\end{align*}
(i) above uses the fact that $e(x), \hat{e}_{N}(x)\in [0, 1]$ and (ii)
uses assumption \textbf{A}1 and the condition that
$\E [1 / e(X)] < \infty$.

\subsection{Proof of Theorem \ref{thm:int_conformal}}
Let $n = |\Z_{\ca}|$ and $(X_{n+1}, C_{n+1})$ be an independent
copy of $(X, C)$. Further let
\[V_{n+1} = \max\{\hat{m}^{L}(X_{n+1}; \Z_{\tr}) - C_{n+1}^{L}, C_{n+1}^{R} - \hat{m}^{R}(X_{n+1}; \Z_{\tr})\}.\]
Conditional on $\Z_{\tr}$, $V_{1}, \ldots, V_{n}, V_{n + 1}$ are exchangeable. Then
\[\P\lb V_{n+1}\le \qt\lb 1 - \gamma; \frac{1}{n+1}\sum_{i=1}^{n+1}\delta_{V_{i}}\rb\rb\ge 1 - \gamma.\]
By definition,
\[\qt\lb 1 - \gamma; \frac{1}{n+1}\sum_{i=1}^{n+1}\delta_{V_{i}}\rb = \qt\lb (1 - \gamma)\frac{n + 1}{n}; \frac{1}{n}\sum_{i=1}^{n}\delta_{V_{i}}\rb = \eta.\]
As a consequence,
\[\P\lb C_{n+1}\in \hat{\cC}(X_{n+1})\rb = \P\lb V_{n+1}\le \eta\rb\ge 1 - \gamma.\]

\section{Additional Experimental Results}\label{app:expr}

\begin{figure}[hbp]
  \centering
  \includegraphics[width = 0.9\textwidth]{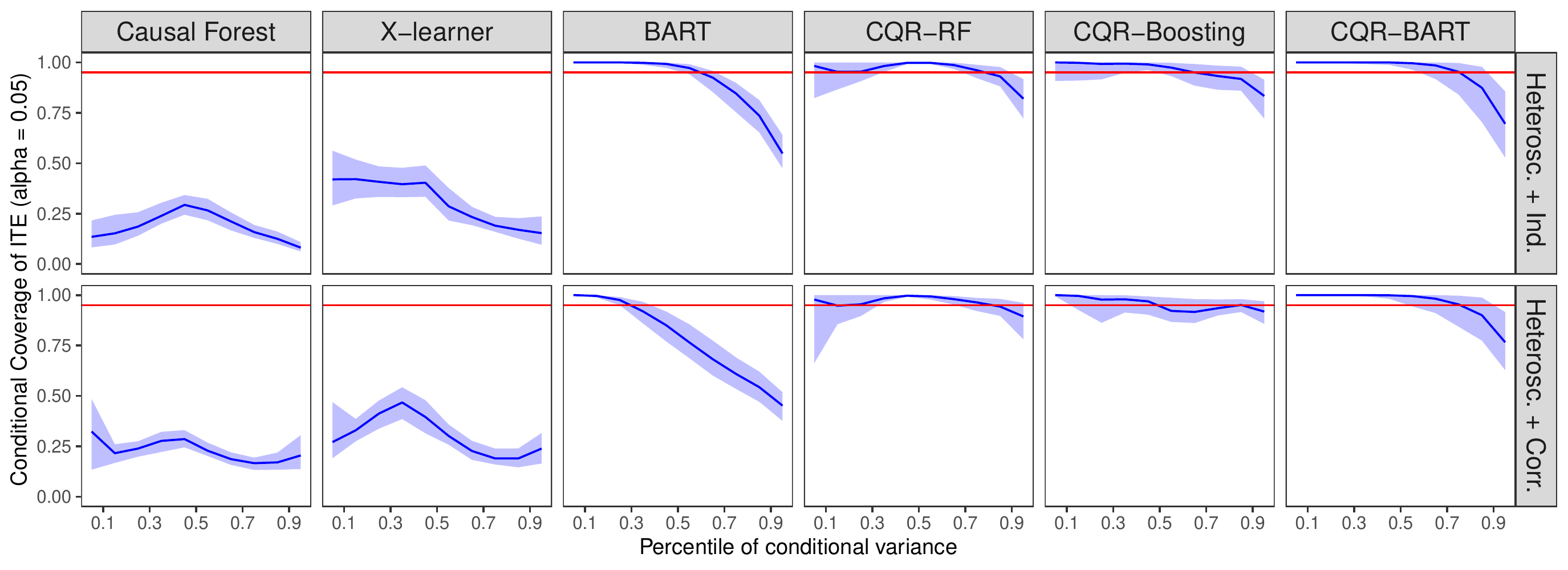}
  \caption{Estimated conditional coverage of ITE as a function of the
    conditional variance $\sigma^{2}(x)$ for the heteroscedastic cases
    from Section \ref{subsec:simul}. Here, $d = 100$ and
    $\alpha = 0.05$. The blue curves correspond to the median and the 
    boundaries of the blue confidence bands correspond to the $95\%$
    and $5\%$ quantiles of these estimates across $100$ replicates.}
\label{fig:extra1}
\end{figure}

\begin{figure}
  \centering
  \includegraphics[width = 0.9\textwidth]{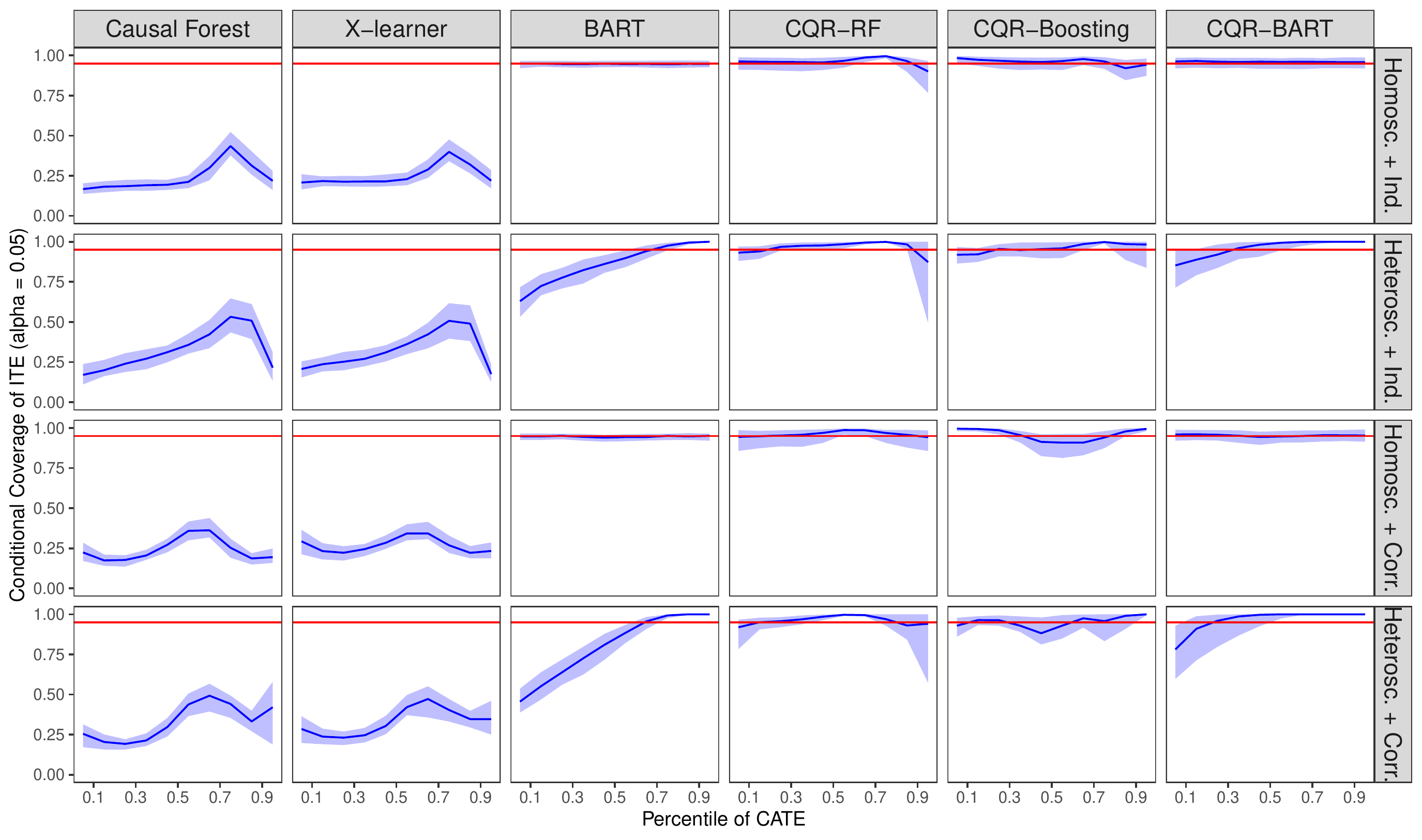}
  \caption{Estimated conditional coverage of ITE as a function of the
    CATE $\tau(x)$ for all scenarios with $d = 10$ from Section \ref{subsec:simul}. Everything else is as in Figure \ref{fig:extra1}.} 
\label{fig:extra2}
\end{figure}

\begin{figure}
  \centering
  \includegraphics[width = 0.9\textwidth]{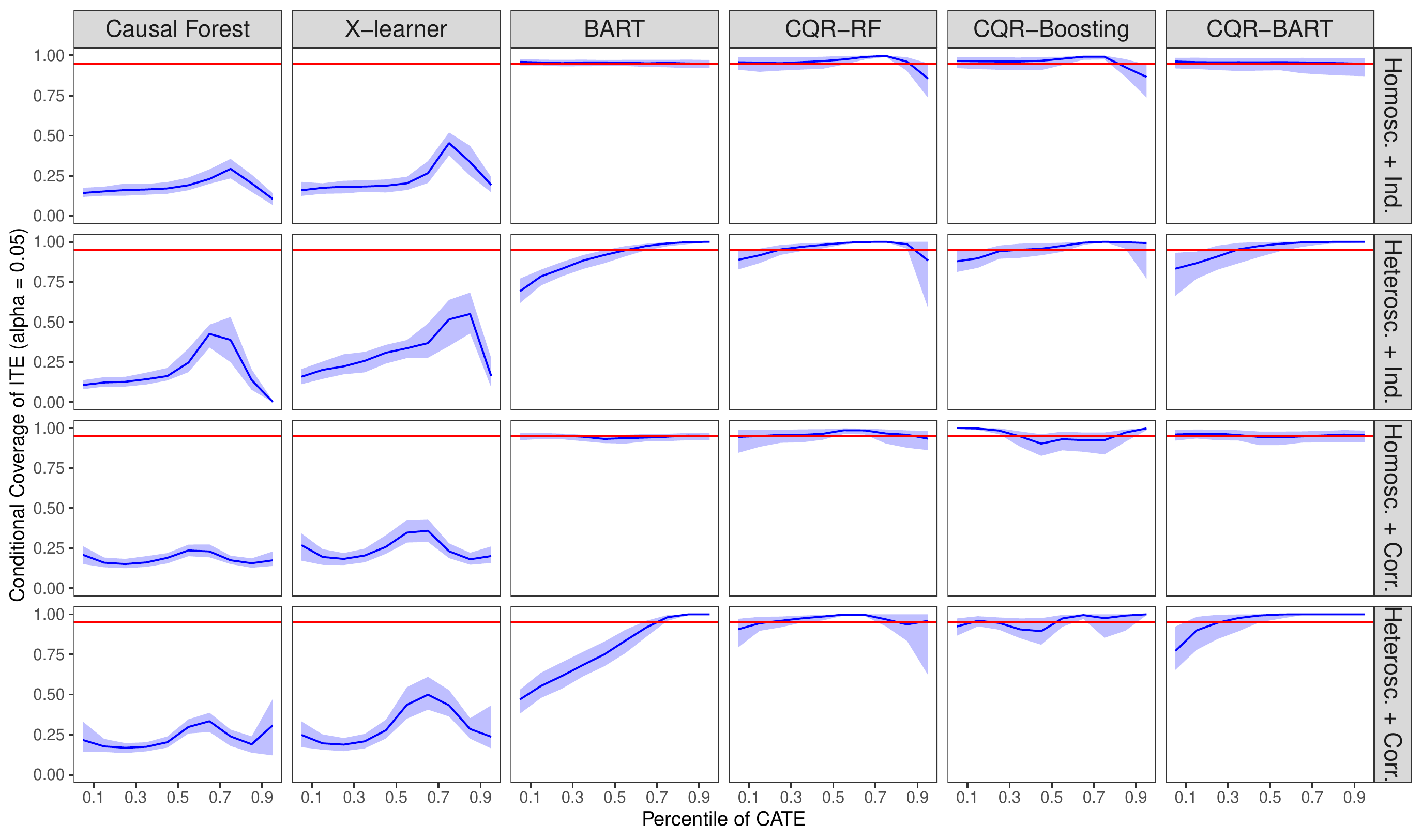}
  \caption{Estimated conditional coverage of ITE as a function of the
    CATE $\tau(x)$ for all scenarios with $d = 100$ from Section
    \ref{subsec:simul}. Everything else is as in Figure
    \ref{fig:extra1}. The difference with Figure \ref{fig:extra2} is
    the value of the dimension $d$. Yet, we can observe a very similar
    behavior.}
\label{fig:extra3}
\end{figure}


\end{document}